\documentclass[5p,longtitle, number, sort&compress]{elsarticle}

\pdfoutput=1
\usepackage[utf8]{inputenc}
\usepackage{lineno}
\usepackage{import}
\usepackage{siunitx}
\usepackage{amssymb}
\usepackage{multirow}
\usepackage[italic]{heppennames}

\usepackage[dvipsnames]{xcolor} 

\RequirePackage{xspace}

\RequirePackage{hyperref}
\begin{document}

\begin{frontmatter}

\title{The LUX-ZEPLIN (LZ) Experiment}

% 1 
\author[1,2]{D.S.~Akerib}
% 2 
\author[3]{C.W.~Akerlof}
% 3 
\author[4]{D.~Yu.~Akimov}
% 4 
\author[5]{A.~Alquahtani}
% 5 
\author[6]{S.K.~Alsum}
% 6 
\author[1,2]{T.J.~Anderson}
% 7 
\author[7]{N.~Angelides}
% 8 
\author[8]{H.M.~Ara\'{u}jo}
\author[6]{A.~Arbuckle}
% 9 
\author[9]{J.E.~Armstrong}
% 10 
% ...

\address[1]{SLAC National Accelerator Laboratory, Menlo Park, CA 94025-7015, USA}

\address[2]{Kavli Institute for Particle Astrophysics and Cosmology, Stanford University, Stanford, CA  94305-4085 USA}

\address[3]{University of Michigan, Randall Laboratory of Physics, Ann Arbor, MI 48109-1040, USA}

\address[4]{National Research Nuclear University MEPhI (NRNU MEPhI), Moscow, 115409, RUS}

\address[5]{Brown University, Department of Physics, Providence, RI 02912-9037, USA}

\address[6]{University of Wisconsin-Madison, Department of Physics, Madison, WI 53706-1390, USA}

\address[7]{University College London (UCL), Department of Physics and Astronomy, London WC1E 6BT, UK}

\address[8]{Imperial College London, Physics Department, Blackett Laboratory, London SW7 2AZ, UK}

\address[9]{University of Maryland, Department of Physics, College Park, MD 20742-4111, USA}

% ...

\begin{abstract}

We describe the design and assembly 
of the LUX-ZEPLIN experiment, a 
direct detection search for cosmic WIMP dark matter particles.
The centerpiece of the experiment is a large liquid xenon
time projection chamber sensitive
to low energy nuclear recoils. Rejection of 
backgrounds is enhanced by a Xe skin veto detector and 
by a liquid scintillator Outer Detector 
loaded with gadolinium for efficient neutron capture and tagging.
LZ is located in the Davis Cavern at the 4850' level of
the Sanford Underground Research Facility in Lead, 
South Dakota, USA.
We describe the major subsystems of the experiment 
and its key design features and requirements. 

\end{abstract}

\end{frontmatter}

%\linenumbers

\section{Overview}
\label{sec:overview}

In this article we describe the design and 
assembly of the LUX-ZEPLIN (LZ) experiment, a 
search for dark matter particles at the 
Sanford Underground Research Facility (SURF) in Lead, 
South Dakota, USA. LZ is capable of observing 
low energy nuclear recoils, the 
characteristic signature of the scattering 
of WIMPs (Weakly Interacting Massive Particles).
It is hosted in the Davis Campus 
water tank at SURF, formerly the home of the LUX  
experiment\cite{akerib:2012ys}. LZ features a 
large liquid xenon (LXe) time projection 
chamber (TPC), a well-established technology
for the direct detection of WIMP dark matter 
for masses greater than a few GeV.
The detector design and experimental 
strategy derive strongly from the 
LUX and ZEPLIN--III experiments~\cite{Akerib:2016vxi,akimov:2011tj}. 
A Conceptual Design Report and a Technical Design Report were completed in 
2015 and 2017, respectively~\cite{Mount:2017qzi,akerib:2015cja}. 
The projected cross-section 
sensitivity of the experiment is $1.5\times 
10^{-48}$ cm$^2$ for a 40 GeV/c$^2$ WIMP (90\%~C.L.)~\cite{Akerib:2018lyp}.

\begin{figure*}[t!]
\centering
\includegraphics[trim={0 0.0cm 0 0.0cm},clip,width=0.65\linewidth]{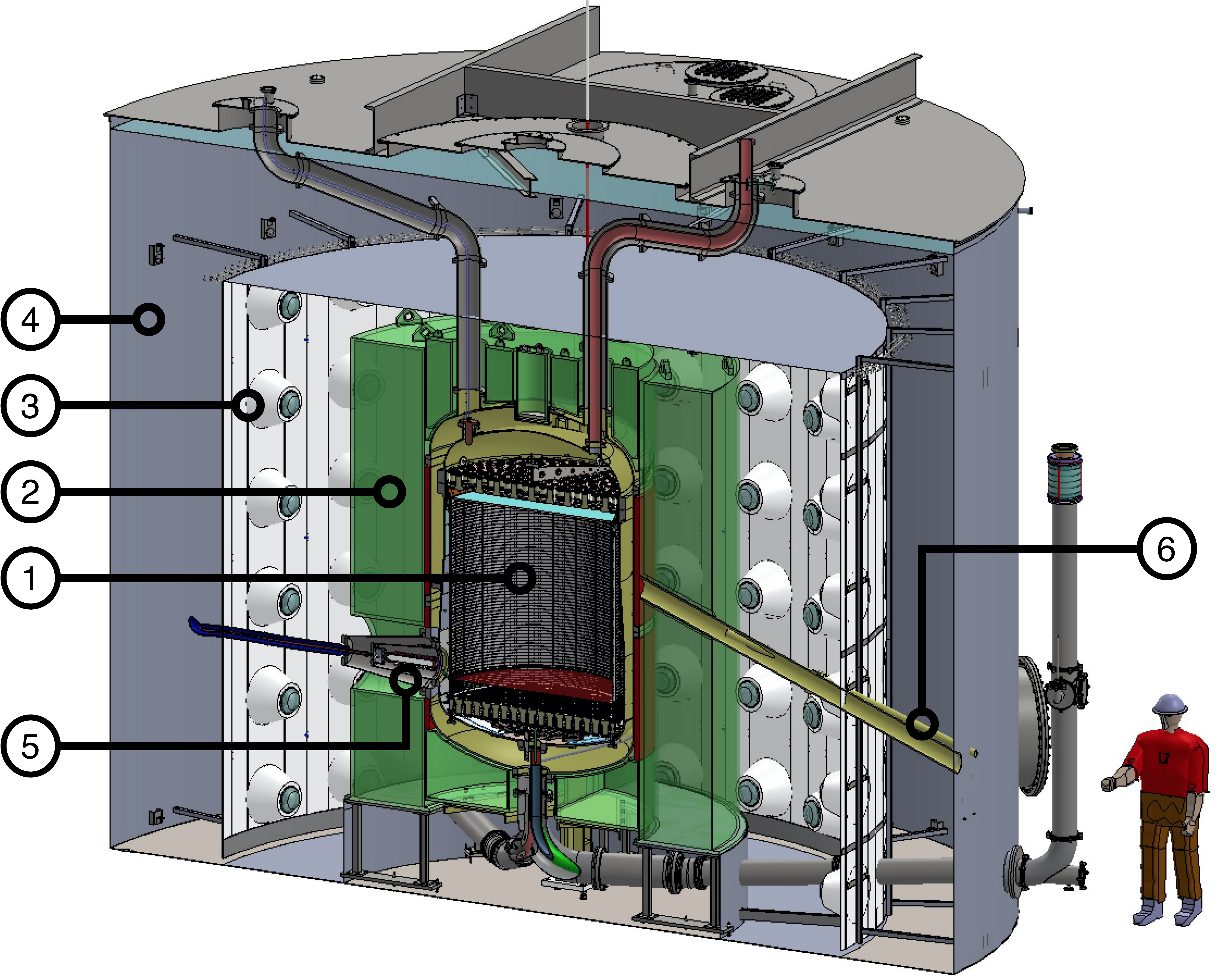}
\caption{Rendering of the LZ experiment, showing the major detector subsystems. At the center is the liquid xenon TPC (1), monitored by two arrays of PMTs and serviced by various cable and fluid conduits (upper and lower). The TPC is contained in a double-walled vacuum insulated titanium cryostat and surrounded on all sides by a GdLS Outer Detector (2). The cathode high voltage connection is made horizontally at the lower left (5). The GdLS is observed by 
a suite of 8" PMTs (3) standing in the water (4) which provides shielding for the detector. The pitched conduit on the right (6) allows for neutron calibration sources to illuminate the detector.}
\label{fig:LZSolid} 
\end{figure*}

A cutaway drawing of the experiment is shown in Fig.~\ref{fig:LZSolid}.  The LZ TPC monitors 7 active tonnes (5.6 tonnes fiducial) of LXe above its cathode. Ionizing interactions in the active region create prompt and secondary scintillation signals (‘S1’ and ‘S2’), and these are observed as photo-electrons (PEs) by two arrays of photomultiplier tubes (PMTs). The nature of the interaction, whether electronic recoil (‘ER’) or nuclear recoil (‘NR’), is inferred from the energy partition between S1 and S2. The location of the event is measured from the drift time delay between S1 and S2 (z coordinate) and from the S2 spatial distribution (x and y coordinates). The TPC is housed in an inner cryostat vessel (ICV), with a layer of ‘skin’ LXe acting as a high voltage stand-off. The skin is separately instrumented with PMTs to veto gamma and neutron interactions in this region. The ICV is suspended inside the outer cryostat vessel (OCV), cooled by a set of LN thermosyphons, and thermally isolated by an insulating vacuum. Both the ICV and OCV are fabricated from low radioactivity titanium~\cite{Akerib:2017iwt}. The cryostat stands inside the Davis Campus water tank, which provides shielding from laboratory gammas and neutrons. An additional set of PMTs immersed in the water observe an Outer Detector (OD) comprised of acrylic vessels (AVs) surrounding the cryostat. The AVs contain organic liquid scintillator loaded with Gadolinium (GdLS) for efficient neutron and gamma tagging. The water tank and OD are penetrated by various TPC services, including vacuum insulated conduits for LXe circulation, instrumentation cabling, neutron calibration guide tubes, and the cathode high voltage (HV) connection.
  
One goal of the experimental architecture is to minimize the amount of underground fabrication and assembly of the various detector sub-systems. The LZ TPC is assembled and integrated into the ICV in a surface laboratory cleanroom at SURF, with the ICV outer diameter taking maximal advantage of the available space in the Yates shaft. The OCV and OD, being larger than the ICV, cannot be transported underground in monolithic form. Therefore the OCV is segmented into three flanged components and integrated and sealed in the Davis Campus water tank, while the OD is subdivided into ten hermetic AVs. This architecture does not require any underground titanium welding or acrylic bonding.
 
Besides the instrumented skin and OD, several other design choices distinguish LZ from its LUX predecessor. The cathode HV connection, for example, is made at a side port on the cryostat, while the PMT cables from the lower half of the TPC are pulled from the bottom. The heat exchanger for LXe condensation, evaporation, and circulation is located in a separate and dedicated cryostat outside the water tank, with LXe being circulated to and from the bottom of the detector through vacuum insulated transfer lines. To continuously reject heat from the LN thermosyphon systems, a cryocooler is installed above the water tank in the Davis Campus. This eliminates the need to transport LN to the underground, except during cryocooler maintenance and repair.
 
The experimental strategy is driven by the need to control radon, krypton, and neutron backgrounds. Control of dust on all xenon-wetted parts is essential, since it can be an important source of radon. Kr is removed from the vendor supplied xenon using an off-site charcoal chromatography facility. This purification step takes place prior to the start of underground science operations.  Gamma backgrounds are highly suppressed by the self-shielding of the TPC, and by careful control and selection of detector materials. Neutrons from spontaneous fission and alpha capture on light nuclei are efficiently tagged and vetoed by the OD and skin.

\section{The Xenon Detector: TPC and Skin}

\label{sec:TPC}

\begin{figure}[h!]
\centering
\includegraphics[trim={0 0.0cm 0 0.0cm},clip,width=0.80\linewidth]{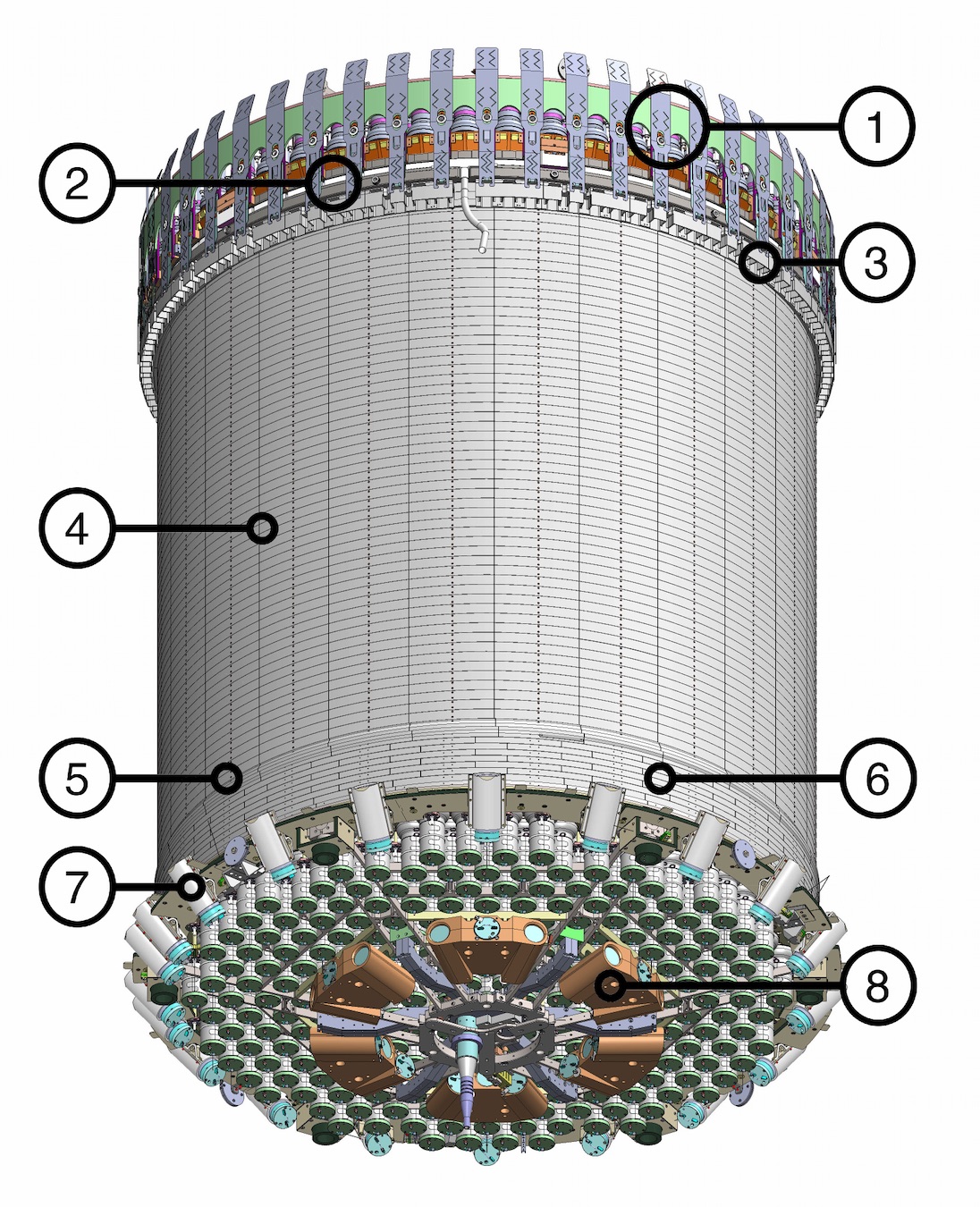} \\
\vspace{0.5cm}
\includegraphics[trim={0 0.0cm 0 0.0cm},clip,width=0.65\linewidth]{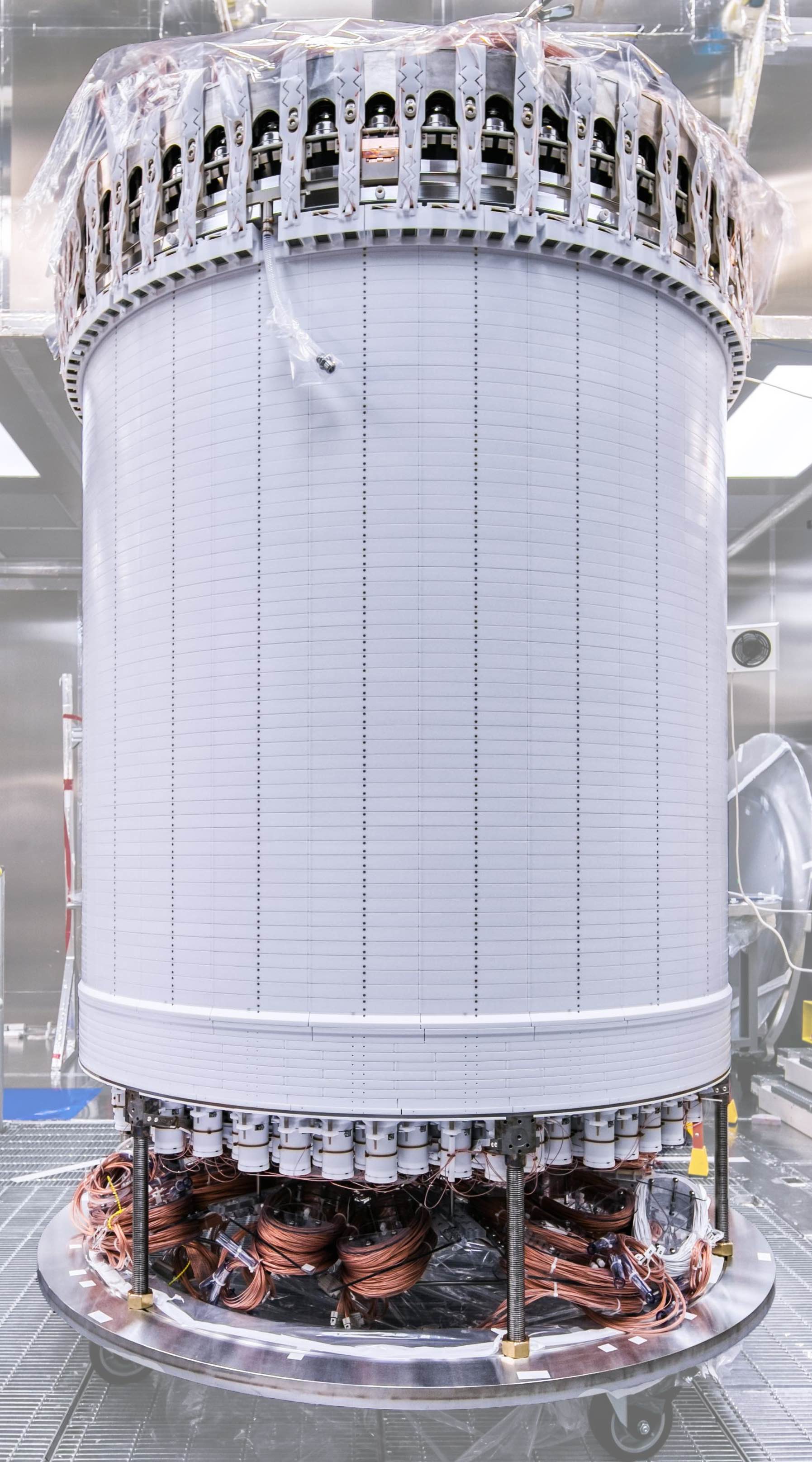}
\caption{The assembled Xenon Detector. Upper panel labels: 1-Top PMT array; 2-Gate-anode and weir region (liquid level); 3-Side skin PMTs (1-inch); 4-Field cage; 5-Cathode ring; 6-Reverse field region; 7-Lower side skin PMTs (2-inch); 8-Dome skin PMTs (2-inch). Lower panel photo by Matthew Kapust, Sanford Underground Research Facility.}
\label{fig:TPC-overview} 
\end{figure}

The Xenon Detector is composed of the TPC and its Xe Skin Veto companion. The central TPC contains 7 tonnes of active LXe which constitutes the WIMP target. This volume measures approximately 1.5~m in diameter and height, and is viewed by two arrays of PMTs. The liquid phase produces prompt S1 pulses. This is topped by a thin layer of vapor (8 mm thick) where delayed S2 electroluminescence light is produced from ionization emitted across the surface. Around and underneath the TPC, the Xe Skin detector contains an additional $\sim$2~tonnes of liquid, also instrumented with PMTs. This space is required for dielectric insulation of the TPC but it constitutes an anti-coincidence scintillation detector in its own right. An overview of the Xenon Detector is shown in Fig.~\ref{fig:TPC-overview}.

The design of the Xenon Detector optimizes: i) the detection of VUV photons generated by both S1 and S2, through carefully chosen optical materials and sensors both in the TPC and the Xe Skin; and ii) the detection of ionization electrons leading to the S2 response, through carefully designed electric fields in the various regions of the TPC. The hardware components involved in the transport and detection of photons and of electrons in the detector are described in Sections~\ref{sec:opticaltpc} and \ref{sec:electricaltpc}. Section~\ref{sec:fluidstpc} describes the flow and the monitoring of the LXe fluid itself.

\subsection{Optical Performance of the TPC}
\label{sec:opticaltpc}

Both the S1 and the S2 signals produced by particle interactions consist of vacuum ultraviolet (VUV) photons produced in the liquid and gas phases, respectively. It is imperative to optimize the detection of these optical signals. For the S1 response, the goal is to collect as many VUV photons as possible, as this determines the threshold of the detector. This is achieved primarily by the use of high quantum efficiency (QE) PMTs optimized for this wavelength region, viewing a high-reflectance chamber covered in PTFE, and by minimizing sources of photon extinction in all materials. Good photocathode coverage is also essential. For the S2 response, the gain of the electroluminescence process makes it easier to collect enough photons even at the lowest energies, and the main design driver is instead to optimize the spatial resolution, especially for peripheral interactions.

The TPC PMTs are 3--inch diameter Hamamatsu R11410--22, developed for operation in the cold liquid xenon and detection of the VUV luminescence. The “-22” variant was tuned for LZ in particular: both for ultra-low radioactivity and for resilience against spurious light emission observed at low temperature in previous variants~\cite{Akimov:2015}. The average cold QE is 30.9\% after accounting for the dual photoelectron emission effect measured for xenon scintillation~\cite{Paredes:2018hxp}. Key parameters were tested at low temperature for all tubes, including pressure and bias voltage resilience, gain and single photoelectron response quality, afterpulsing and dark counts. These are critical parameters that directly influence the overall performance of the detector. The procurement, radioassay and performance test campaign lasted for nearly three years. The PMTs are powered by resistive voltage divider circuits attached to the tubes inside the detector. The voltage ladder is that recommended by Hamamatsu, using negative bias to extract the signal near ground potential (two independent cables are used for signal and bias). The nominal operating gain of the PMTs is $3.5\times 10^6$ measured at the end of the signal cables.

\begin{figure}[h!]
\centering
\includegraphics[trim={0 0.0cm 0 0.0cm},clip,width=0.85\linewidth]{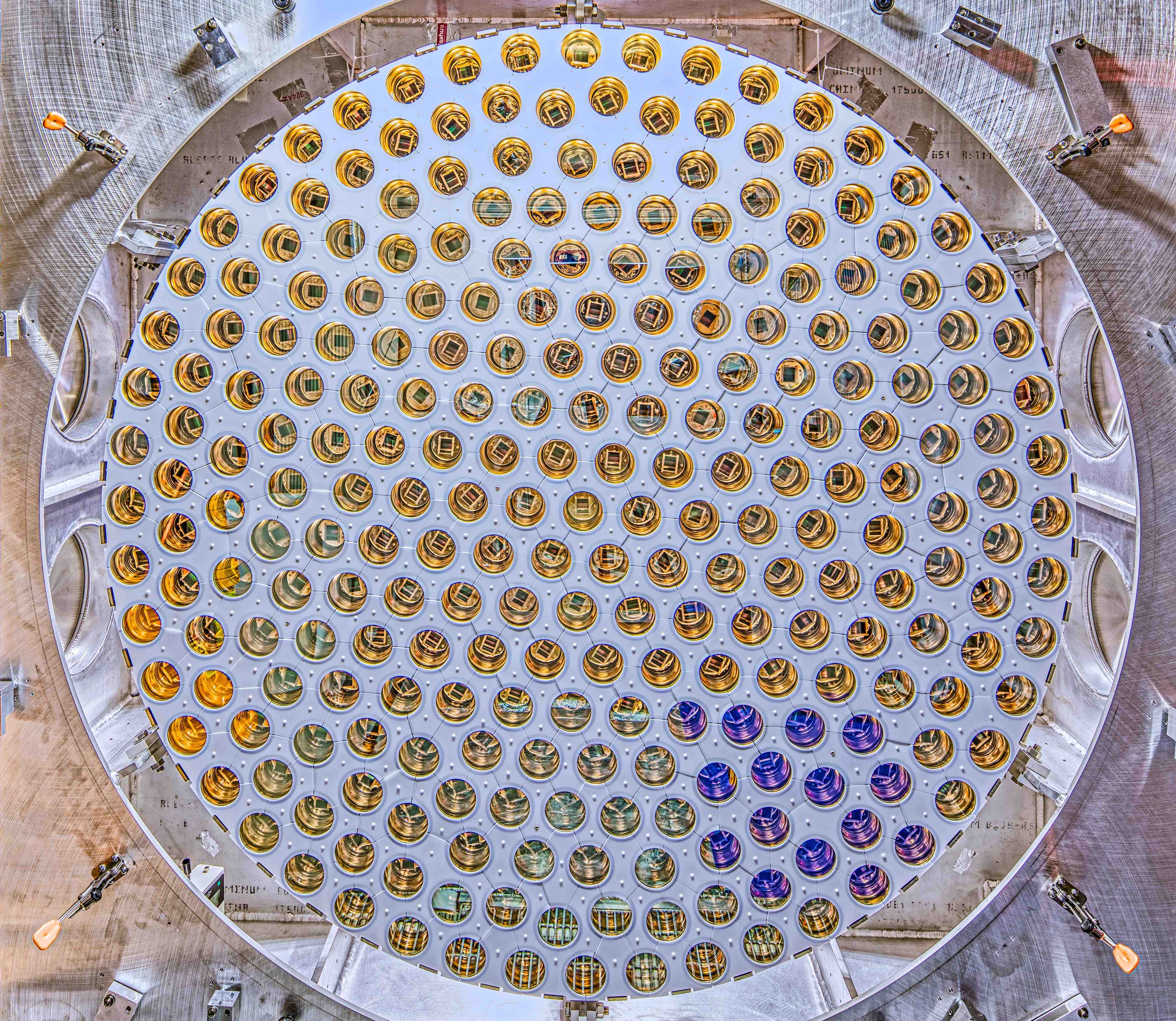} \\
\vspace{0.5cm}
\includegraphics[trim={0 0.0cm 0 0.0cm},clip,width=0.85\linewidth]{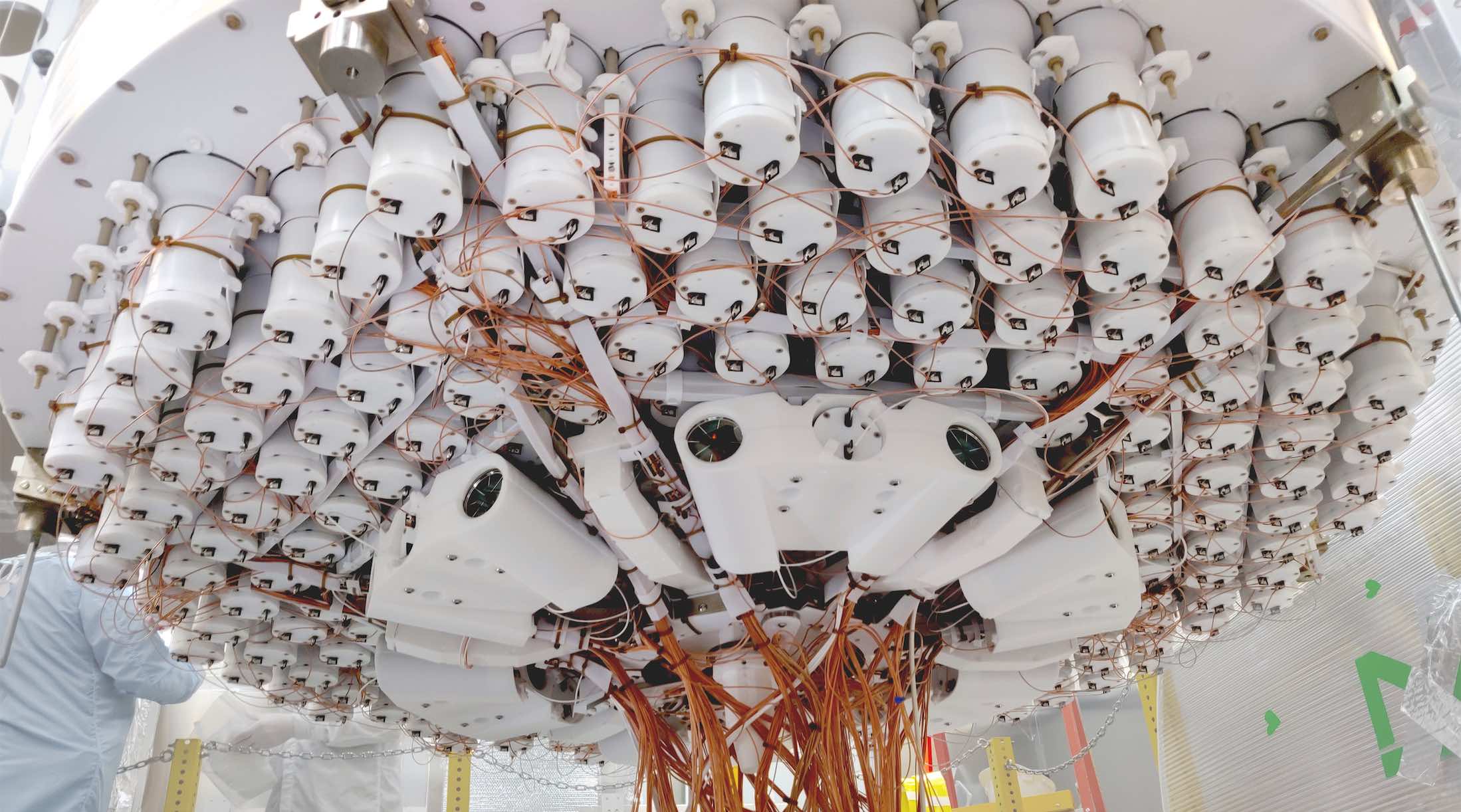}
\caption{Arrays of R11410--22 PMTs viewing the TPC. Upper panel: front view of the top PMT array within its assembly and transportation enclosure. Note the circular PMT arrangement at the periphery transitioning to compact hexagonal towards the center, and the coverage of non-sensitive surfaces by interlocking pieces of highly reflective PTFE. Photo by Matthew Kapust, Sanford Underground Research Facility. Lower panel: Back view of bottom array PMTs in hexagonal arrangement, showing cable connections and routing as well as 18 2-inch dome PMTs, which are part of the skin veto system. Also visible are the titanium support trusses and the LXe distribution lines.  Most surfaces are covered in PTFE to aid light collection.}
\label{fig:PMT-arrays} 
\end{figure}

Two PMT arrays detect the xenon luminescence generated in the TPC. These are shown in Fig.~\ref{fig:PMT-arrays}. An upward-looking “bottom” array immersed in the liquid contains 241 units arranged in close-packed hexagonal pattern. A downward-looking “top” array located in the gas phase features 253~units arranged in a hybrid pattern that transitions from hexagonal near the center to circular at the perimeter. This design was chosen to optimize the position reconstruction of the S2 signal for interactions near the TPC walls, a leading source of background in these detectors. The structural elements of the arrays are made from low-background titanium. These include a thin plate reinforced by truss structures with circular cut-outs to which the individual PMTs are attached. In the bottom array this plate sits at the level of the PMT windows. The exposed titanium is covered with interlocking PTFE pieces to maximize VUV reflectance. The PMTs are held by Kovar belts near their mid-point and attached to this plate by thin PEEK rods. In the top array the structural plate is located at the back of the tubes, and the gaps between PMT windows are covered with more complex interlocking PTFE pieces secured to the back plate. The array designs ensure that mechanical stresses induced by the thermal contraction of PTFE and other materials does not propagate significantly to the PMT envelopes. A number of blue LEDs (Everlight 264-7SUBC/C470/S400) are installed behind plastic diffusers between PMTs at the face of both arrays. These are used to help optimize and calibrate the PMT gains and the timing response of the detector. The assembly and transport of the PMT arrays required a robust QA process to prevent mechanical damage, dust contamination, and radon-daughter plate-out. At the center of this program were specially-designed hermetic enclosures that protected the arrays during assembly, checkout, transport and storage until assembly into the TPC at SURF. 

A key element of the optical systems is the $\sim$20~km of cabling used for PMT and sensor readout. A 50-Ohm coaxial cable from Axon Cable S.A.S. (part no. P568914A\textasciicircum) was selected for electrical and radioactivity performance. This cable has a copper-made inner conductor and braid, and extruded FEP insulator and outer sleeve (1.3-mm nominal diameter). The 12~m span from the detector to the external feed-throughs means that signal attenuation, dispersion and cross-talk are important considerations. The individual cables were pre-assembled into bundles which are routed together through two conduits that carry the cables from the top and bottom of the detector. An additional consideration is the potential for radon emanation. This is especially important for the fraction of the cabling located near room temperature. Low intrinsic radioactivity of the cable materials can be easily achieved, but dust and other types of contamination trapped within the fine braid during manufacture can be problematic. We developed additional cleanliness measures with Axon to mitigate this and have opted for a jacketed version which acts as a further radon barrier. In addition, the xenon flow purging the cable conduits is directed to the inline radon-removal system described in Sec.~\ref{sec:XeHandling}.

After the PMTs, the next main optical component of the detector is the PTFE that defines the field cage and covers the non-sensitive detector components. The optical performance of the detector depends strongly on its VUV reflectivity -- VUV photons reflect several times off PTFE surfaces before detection -- and the radiopurity of this material is also critical due to its proximity to the active volume. We identified both the PTFE powder and process that optimized radiological purity and VUV reflectivity in liquid xenon during a long R\&D campaign~\cite{Neves:2016tcw,Haefner:2016ncn}. The PTFE selected was Technetics 8764 (Daiken M17 powder), whose reflectivity when immersed in LXe we measured as 0.973 ($>$0.971 at 95\% C.L.). Our data are best fitted by a diffuse-plus-specular reflection model for this particular material, which was tested using the procedures described in Ref.~\cite{Neves:2016tcw}. Most PTFE elements were machined from moulds of sintered material, while thinner elements are `skived' from cast cylinders manufactured from the same powder.

Other factors influence the photon detection efficiency (PDE) for S1 and S2 light in the TPC. These include photon absorption by the electrode grids (wires and ring holders) and absorption by impurities in the liquid bulk. With realistic values for these parameters our optical model predicts a photon detection efficiency of around 12\% for S1 light.

The optical design of the S2 signal is optimized for robust reconstruction of low energy events at the edge of the TPC, in particular from the decay of Rn daughters deposited on the field cage wall, termed ``wall events".  These events may suffer charge loss, thus mimicking nuclear recoils~\cite{lee2015}.  If mis-reconstructed further into the active region, they can be a significant background. Our aim is to achieve $\sim$10$^6$:1 rejection for this event topology in the fiducial volume. A detailed study of this issue led to the adoption of a circular PMT layout near the detector edge, with the final PMT row overhanging the field cage inner walls, an optimized distance between the top PMT array and the liquid surface, and an absorbing optical layer (Kapton foil) covering the lateral conical wall in the gas phase. 

\subsection{The Xe Skin Detector}

An important component of the Xenon Detector is the Xe Skin, the region containing around 2 tonnes of LXe between the field cage and the inner cryostat vessel. A primary motivation for this liquid is to provide dielectric insulation between these two elements. In addition to its electrical standoff function, it is natural to instrument this region for optical readout so that it can act as a scintillation-only veto detector, especially effective for gamma-rays. Also, if the skin were not instrumented, light from particle interactions or electrical breakdown in this region could leak in to the TPC unnoticed and create difficult background topologies. To further suppress this pathology, the LZ field cage is designed to optically isolate the skin from the TPC.

\begin{figure}[h!]
\centering
\includegraphics[trim={0 0.0cm 0 0.0cm},clip,width=0.83\linewidth]{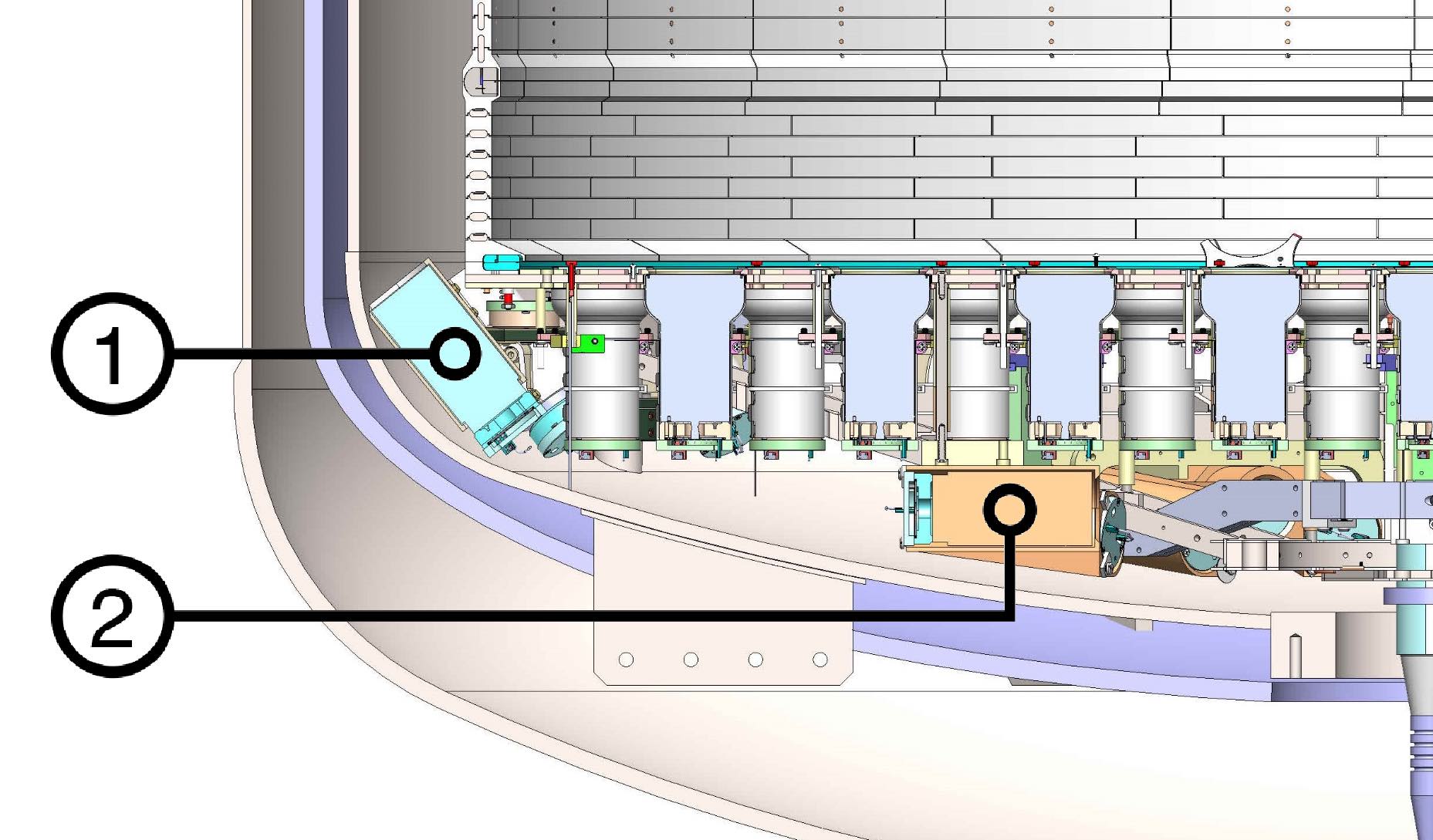} \\
\vspace{0.5cm}
\includegraphics[trim={0 0.0cm 0 0.0cm},clip,width=0.83\linewidth]{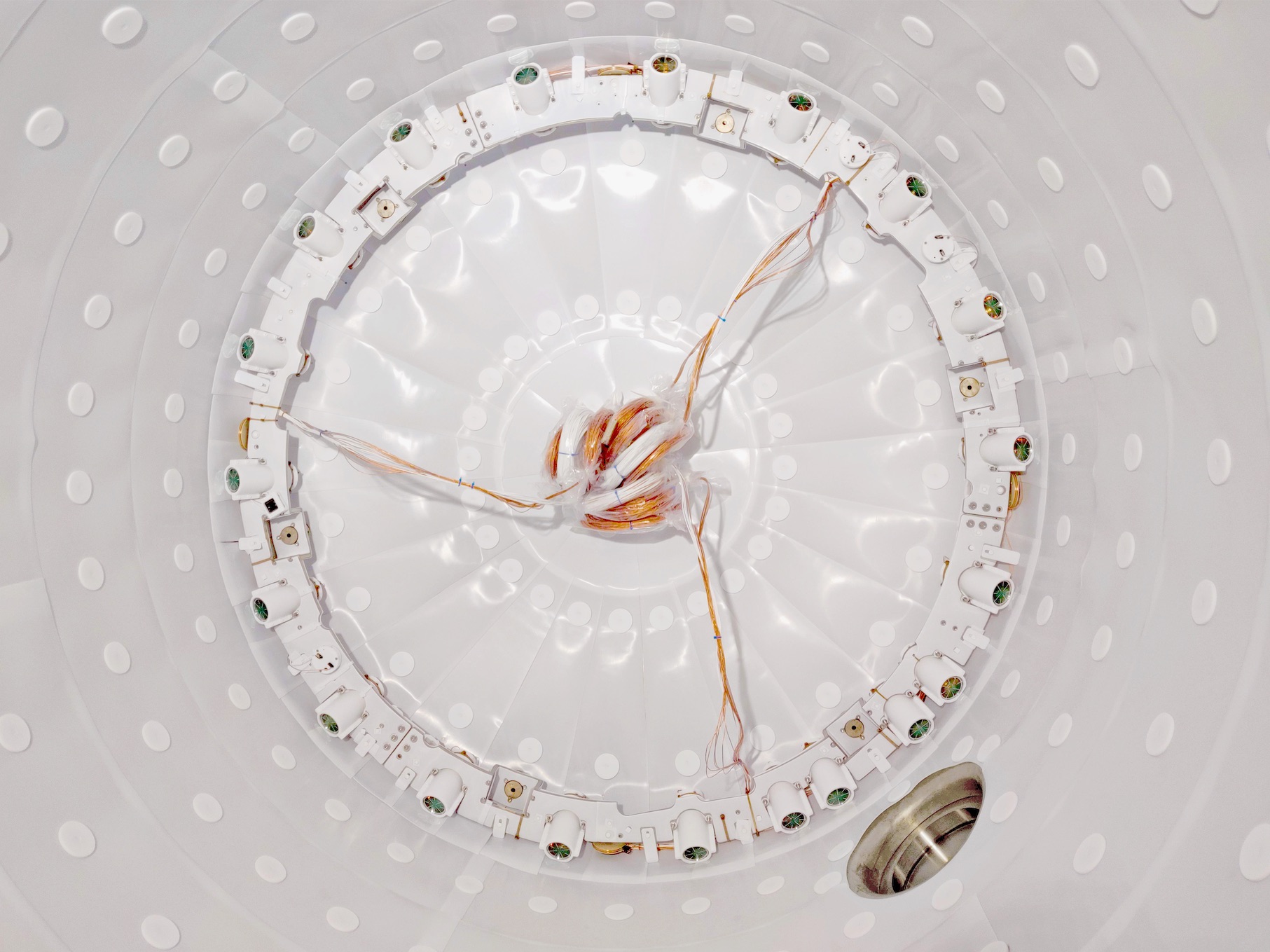}
\caption{Upper panel: CAD section of the TPC below the cathode showing the location of the 2" bottom side skin (1) and lower dome (2) PMTs. Lower panel: Photograph showing the PTFE panelling attached to the ICV that ensures high reflectance in the skin region and the lower side skin PMT ring at the bottom of the vessel.}
\label{fig:SkinDome} 
\end{figure}

The side region of the skin contains 4~cm of LXe at the top, widening to 8~cm at cathode level for increased standoff distance. This is viewed from above by 93 1--inch Hamamatsu R8520-406 PMTs. These are retained within PTFE structures attached to the external side of the field cage, located below the liquid surface. At the bottom of the detector a ring structure attached to the vessel contains a further 20 2--inch Hamamatsu R8778 PMTs viewing upward into this lateral region, as shown in Fig.~\ref{fig:SkinDome}.

The dome region of the skin at the bottom of the detector is instrumented with an additional 18 2--inch R8778 PMTs. These are mounted horizontally below the bottom array, with 12 looking radially outward and 6 radially inward. To enhance light collection, all PMTs in that region and array truss structures are dressed in PTFE. Moreover, PTFE tiles line the ICV sides and bottom dome. To attach the PTFE lining, low profile titanium buttons designed to minimize field effects were epoxied to the ICV wall with MasterBond EP29LPSP cryogenic epoxy. Holes were machined into the PTFE tiles to fit around the buttons, and PTFE washers were attached to the buttons with PTFE screws to secure the tiles in place. These are visible in Fig.~\ref{fig:SkinDome}.

\subsection{TPC Electrostatic Design}
\label{sec:electricaltpc}

The S2 signature detected from particle interactions in the liquid xenon comes from the transport of ionization electrons liberated by the recoiling nucleus or electron, and their subsequent emission into the gas phase above the liquid, where the signal is transduced into a second VUV pulse via electroluminescence. Great care is required to ensure that the various electric field regions in the detector achieve this with high efficiency and with low probability for spurious responses. The LZ detector is instrumented as a traditional three-electrode two-phase detector, with cathode and gate wire-grid electrodes establishing a drift field in the bulk of the liquid, and a separate extraction and electroluminescence region established between the gate and an anode grid. The former sits 5~mm below the surface and the latter is 8~mm into the gas. The nominal operating pressure of the detector is 1.8~bara. At nominal fields, each electron emitted into the gas generates $\sim$820 electroluminescence photons.

\begin{figure}[h!]
\centering
\includegraphics[trim={0 0.0cm 0 0.0cm},clip,width=0.95\linewidth]{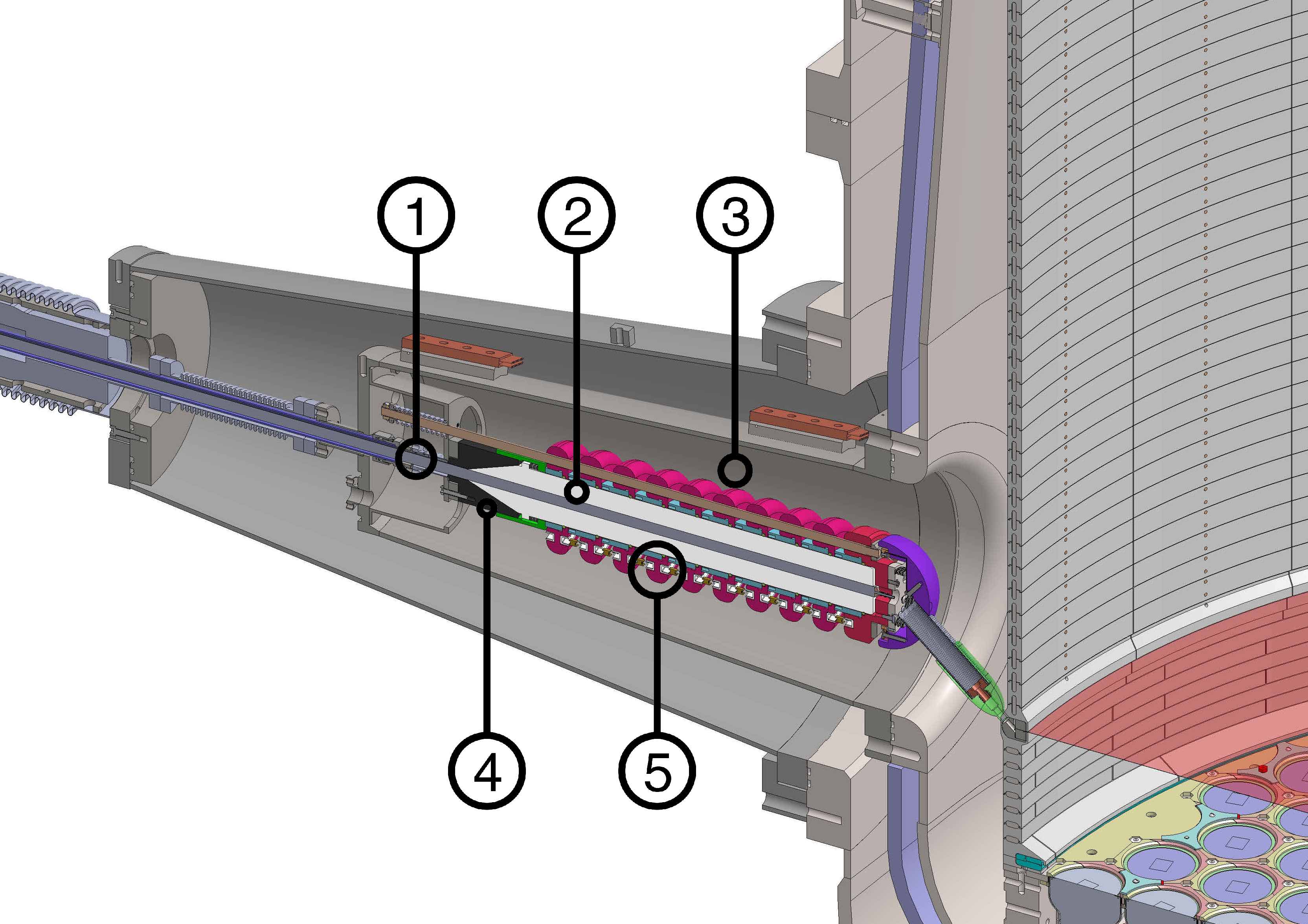}
\caption{The interface of the high voltage system with the cathode. 1-Polyethylene high voltage cable; 2-LXe displacer; 3-LXe space; 4-Stress cone; 5-Grading rings.}
\label{fig:CathodeCone} 
\end{figure}

The nominal 300~V/cm drift field established in the active region of the detector requires application of an operating voltage of $-$50~kV to the cathode grid, which allows LZ to meet its baseline performance for particle discrimination. The design goal is $-$100 kV, the maximum operating voltage for the system. The system to deliver the HV to the cathode grid contains some of the highest fields in the detector. The HV is delivered from the power supply (Spellman SL120N10, 120~kV) via a room-temperature feed-through and into a long vacuum-insulated conduit entering the detector at the level of the cathode grid, as shown in Fig.~\ref{fig:CathodeCone}. Most of the system was tested to $-$120~kV in liquid argon, except for the flexible component connecting the grading structure to the cathode, for which a similarly-shaped part was tested in liquid argon to surface fields 30\% higher than those needed to meet the design goal.

\begin{figure*}[t!]
\centering
\includegraphics[trim={0 0.0cm 0 0.0cm},clip,width=0.95\linewidth]{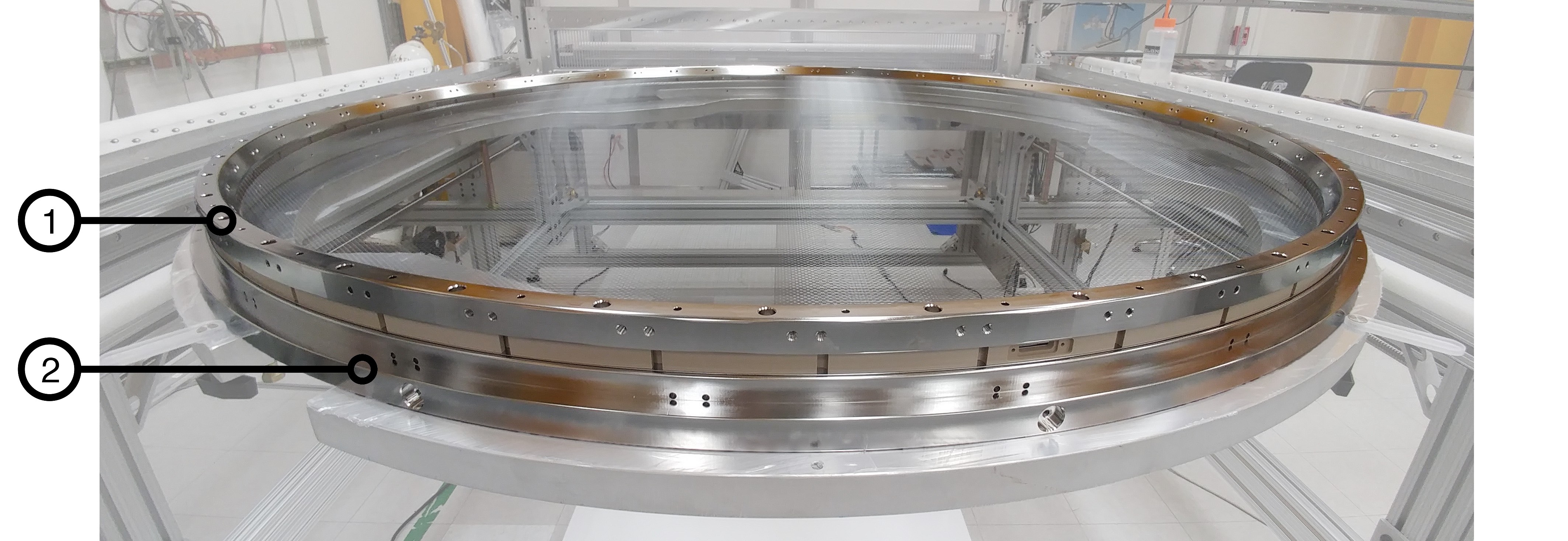}
\caption{The electron extraction region assembly on the loom. 1-Anode grid; 2-Gate grid. During TPC operations the anode is above the LXe surface and the gate is below. The liquid level is registered to this assembly by three weir spillovers.}
\label{fig:FullSizeGrid} 
\end{figure*}

The cable enters the xenon space through two o-rings at the core of the feed-through system mounted on top of the water tank. The space between them is continuously pumped to provide a vacuum guard, monitored by a Residual Gas Analyzer. Located at room temperature and far from the detector, a leak-tight seal to the xenon space can be reliably achieved and the feed-through materials are not a radioactivity concern. Another key feature of the cathode HV system is the reliance on a single span of polyethylene cable (Dielectric Sciences SK160318), connecting the power supply all the way to the cathode in the liquid xenon many meters away. This 150~kV-rated cable features a conductive polyethylene sheath and center core and contains no metal components, avoiding differential contraction and thermal stress issues, and precluding the appearance of insulation gaps between the dielectric and the sheath, which contract equally. The HV line ends in a complex voltage-grading structure near the cathode grid ring where the conductive sheath splays away. This grades the potential along the dielectric, preventing very high field regions inside the detector. The maximum field in the liquid xenon is 35~kV/cm. This voltage grading system ends in a bayonet connector that allows rapid engagement to the cathode ring during installation, minimizing exposure of the detector to radon.

Inside the ICV there is a significant insulation stand-off distance of 4--8 cm between the field cage and the inner vessel, and there is no instrumentation or cabling installed along the length of the skin region, where the field is high and the possibility of discharges and stray light production would be concerning. This region is instead optimized for optical readout, becoming an integral part of the LZ veto strategy.

The drift region is 145.6~cm long between cathode and gate electrodes, and 145.6~cm in diameter, enclosed by a cylindrical field cage which defines the optical environment for scintillation light and shapes the electric field for electron transport. The field cage is constructed of 58 layers of PTFE, which provides insulation and high reflectivity, with a set of embedded titanium electrode rings. The layers are 25~mm tall, the Ti rings 21~mm tall, and each layer of PTFE is azimuthally segmented 24 times. Due to their proximity to the active volume, these are critical materials for LZ. The metal rings are made from the same batch of titanium used for the cryostat~\cite{Akerib:2017iwt} (the PTFE is described above). A key design driver was to achieve a segmented field cage design: to prevent the excessive charge accumulation observed in continuous PTFE panels, and to better cope with the significant thermal contraction of PTFE between ambient and low temperature. The field cage embeds two resistive ladders connecting the metal rings, each with two parallel 1~G$\Omega$ resistors per section (the first step has 1~G$\Omega$ in parallel with 2~G$\Omega$ to tune the field close to the cathode ring). This ladder ensures a vertical field with minimal ripple near the field cage.

The lower PMT array cannot operate near the high potential of the cathode and so a second, more compact ladder is required below that electrode. This reverse field region (RFR) contains only 8 layers with two parallel 5~G$\Omega$ resistors per section, and terminates 13.7~cm away at a bottom electrode grid which shields the input optics of the PMTs from the external field.

The electrode grids are some of the most challenging components of the experiment, both to design and to fabricate.  Mechanically, these are very fragile elements that nonetheless involve significant stresses and require very fine tolerances for wire positioning. Besides optimizing the conflicting requirements of high optical transparency and mechanical strength, electrical resilience was an additional major driver -- spurious electron and/or light emission from such electrodes is a common problem in noble liquid detectors~\cite{rebel:2014uia,Tomas:2018}.

The anode and gate grids are depicted in Fig.~\ref{fig:FullSizeGrid}.
All grids are made from 304 stainless steel ultra-finish wire~\cite{cfw} woven into meshes with a few mm pitch using a custom loom. Key parameters of the four LZ grids are listed in Table~\ref{tab:Grids}. Each wire is tensioned with 250~g weights on both ends, and the mesh is glued onto a holder ring. The glue, MasterBond EP29LPSP cryogenic epoxy, is dispensed by a computer-controlled robotic system. It includes acrylic beads that prevent external stresses from being transferred to the wire crossings. A second metal ring captures the glued region, and the tensioning weights are released after curing. This woven mesh has several advantages over wires stretched in a single direction.  The load on the ring set is azimuthally uniform and purely radial, allowing the mass of the rings to be minimized. The region of non-uniform field near the wires is smaller for a mesh, which improves the uniformity and hence energy resolution obtained in the S2 channel. Finally, a mesh grid has lower field transparency than stretched wires, resulting in a more uniform overall drift field. To preserve high S2 uniformity, it is important that the woven mesh have uniform wire spacing. The loom design included several features to achieve high uniformity during fabrication, and great care was taken in subsequent grid handling to avoid displacing wires.

\begin{table}[tbh]
\setlength{\extrarowheight}{3pt}
\caption[TPC electrode grid parameters]{TPC electrode grid parameters (all \SI{90}{\degree} woven meshes).}
\centering
\begin{tabular} {lrrcc}
\hline
Electrode & Voltage & Diam. & Pitch & Num. \\
 & (kV) & ($\mu$m) & (mm) &  \\
\hline
Anode   &  $+$5.75 & 100 & 2.5& 1169\\
Gate    &  $-$5.75 &  75 & 5.0& 583\\
Cathode &  $-$50.0	 & 100 & 5.0& 579\\
Bottom  &  $-$1.5 &  75 & 5.0& 565\\
\hline
\end{tabular}
\label{tab:Grids} 
\end{table}

\begin{figure}[t!]
\centering
\includegraphics[trim={0 0.0cm 0 0.0cm},clip,width=0.95\linewidth]{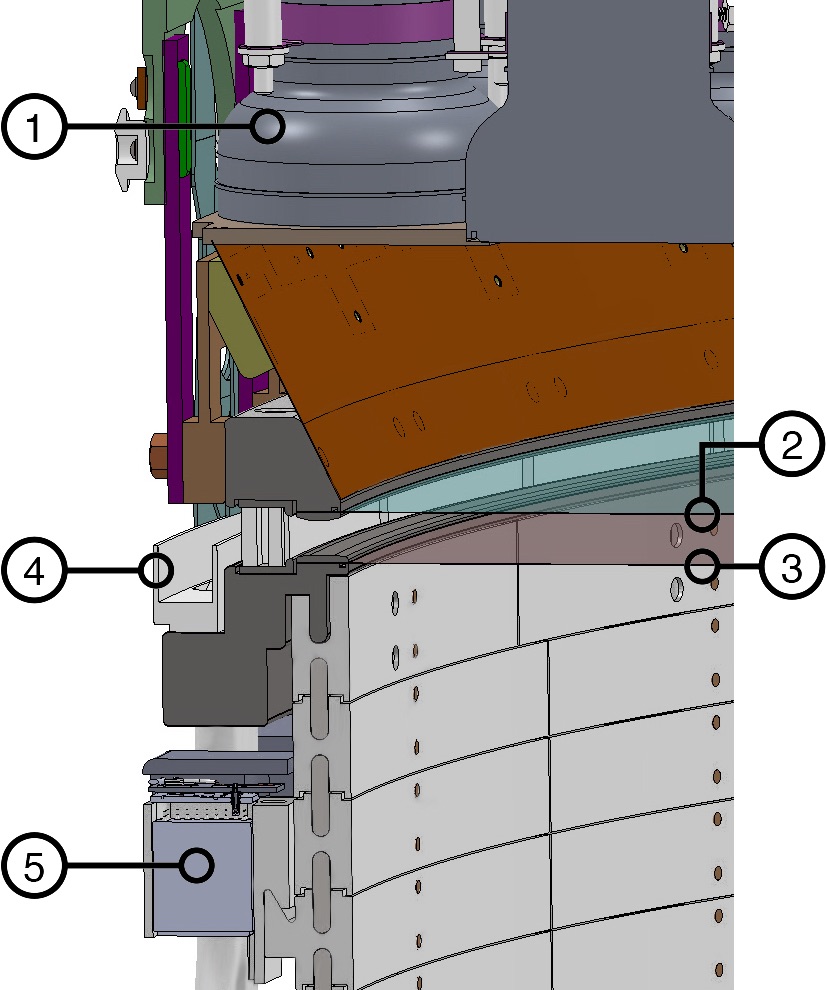}
\caption{The electron extraction and electroluminescence region. 1-TPC PMT; 2-Anode grid; 3-Gate grid; 4-Weir; 5-Xe Skin PMT.}
\label{fig:ER} 
\end{figure}

At the top of the detector, the electron extraction and electroluminescence region, which contains the gate-anode system, is one of its most challenging aspects (see Fig.~\ref{fig:ER}.) It establishes the high fields that extract electrons from the liquid and then produce the S2 light. The quality of the S2 signal is strongly dependent on both the small- and large-scale uniformity achieved in this region. In particular, the anode contains the finest mesh of any LZ grid (2.5~mm pitch) as this drives the S2 resolution. 

An important consideration was the electrostatic deflection of the gate-anode system. We have directly measured the deflection of the final grids as a function of field using a non-contact optical inspection method.  This matches expectations from electrostatic and mechanical modeling, and predicts a $\sim$1.6~mm decrease in the 13~mm gap at the 11.5~kV nominal operating voltage. As a consequence the field in the gas phase varies from 11.5~kV/cm at the center to 10.1~kV/cm at the edge. The combined effect of field increase and gas gap reduction increases the S2 photon yield by 5\% at the center. This effect can be corrected in data analysis. The gate wires sustain the strongest surface fields of any cathodic element in the detector ($\simeq$52 kV/cm, with no grid deflection, and $\simeq$58~kV/cm with 1.6 mm combined gate/anode deflection).

A major QA program was implemented to ensure the high quality of the grids throughout manufacture, cleaning, storage and transport, and to prevent damage and dust contamination. A key feature of this program was a series of measurements of the high voltage behavior in Xe gas of ~1/10th scale and full scale prototype grids, and the final cathode, gate and anode grids. Emission of single electrons to a rate as low as $\sim$Hz was measured via an S2-like electroluminescent signal with PMTs. These measurements confirmed earlier work~\cite{Tomas:2018} showing that electron emission is strongly reduced by passivation, thus the production gate grid was passivated in citric acid for 2 hours at $\sim$125$^{\circ}$F at AstroPak~\cite{Astropak}.

\subsection{Fluid Systems and Sensors}
\label{sec:fluidstpc}

Purified and sub-cooled LXe is prepared by the LXe tower and delivered to the Xenon Detector through two vacuum-insulated supply lines that connect at the ICV bottom center flange (see Sec.~\ref{sec:XeHandling}). One line flushes the lower dome and side skin, the other fans out through a manifold into seven PTFE tubes that penetrate the lower PMT array and supply LXe to the TPC. The fluid returns to the external purification system by spilling over three weirs that establish the liquid surface height.  The weirs have 23.3~cm circumference, are uniformly spaced in azimuth around the top of the TPC, and are mounted to the gate grid ring so that the liquid level is well registered to the location of the gate and anode grids. The weirs drain through three tubes that penetrate the ICV in the side skin region and descend in the insulating vacuum space. The three lines are ganged together near the bottom of the ICV and return liquid to the purification circuit through a common vacuum-insulated transfer line.

A variety of sensors monitor the behavior and performance of the TPC. Six Weir Precision Sensors (WPS) measure the liquid level to within $\approx$20~$\mu$m in the gate-anode region.  An additional WPS is installed in the lower dome to monitor filling and draining of the detector. Long level sensors (LLS) are installed in the LXe tower for providing information during detector filling and for monitoring during normal operation. RF loop antennae (LA) and acoustic sensors (AS) monitor the electrostatic environment of the detector during electrode biasing and thereafter during operation. The combination of WPS and AS sensors will also be used to detect disturbances of the fluid system and especially the liquid surface, such as bubbling or dripping. These will be aided by dedicated resistors installed in the bottom array that will be used to create bubbles in the LXe so that their signature can be characterized. At the top of the detector, a hexapod structure connects the top PMT array to the ICV lid through six displacement sensors, allowing the displacement and tilt between these two elements to be measured to within 0.1~degrees. This is especially important to prevent major stresses arising during cool-down of the TPC. Finally, PT100 thermometers are distributed at both ends of the detector. By design, all sensors and their cabling are excluded from the side skin region and other high electric field regions. All sensors are read out by dedicated electronics attached to flanges enclosed in the signal breakout boxes.

\section{Cryogenics and Xe Handling}

\subsection{Cryostat and cryogenics}
\label{sec:cryostat}

\begin{figure}[h!]
\centering
\includegraphics[trim={0 0.0cm 0 0.0cm},clip,width=0.80\linewidth]{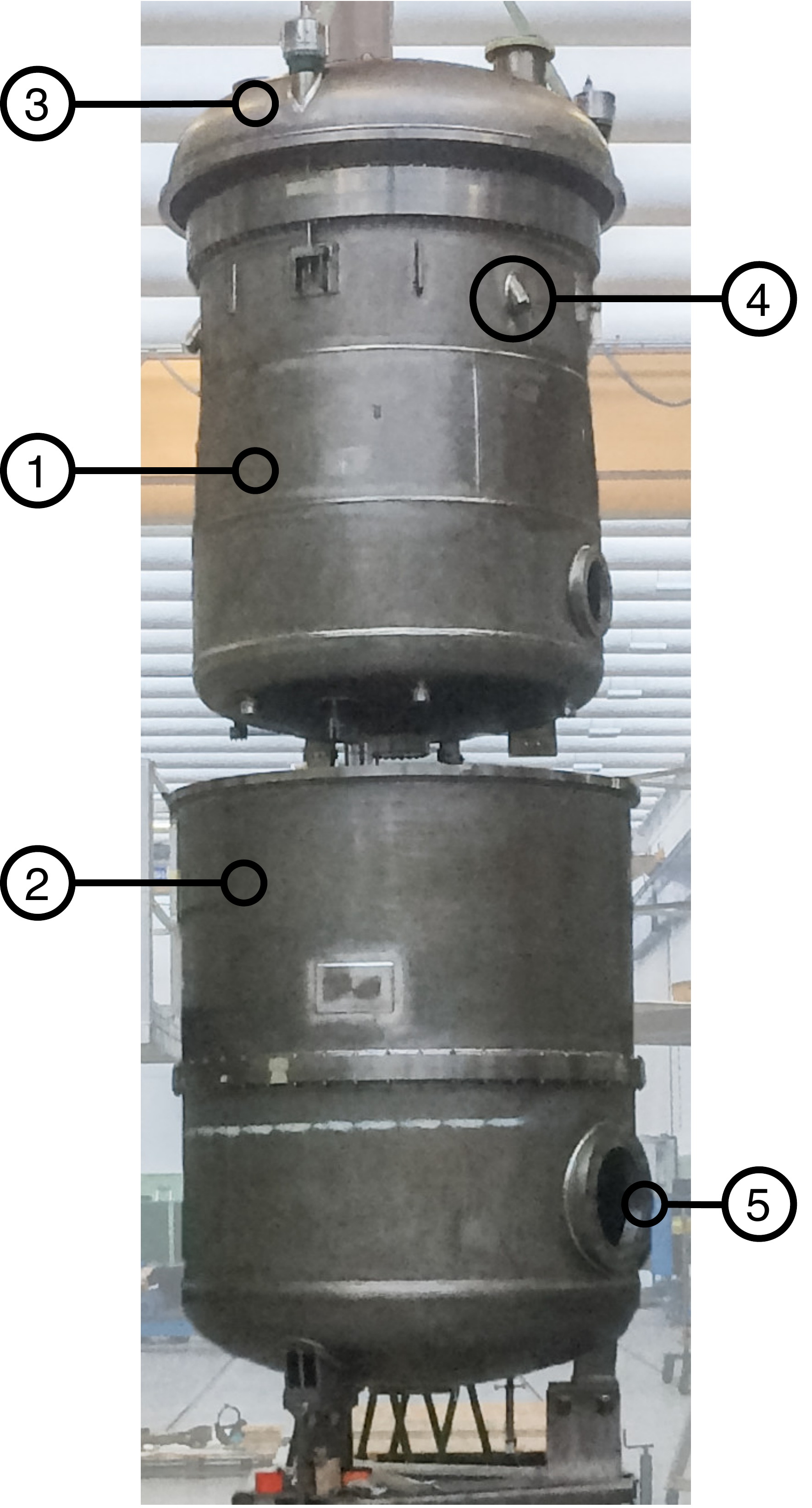}
\caption{The ICV and OCV during a test assembly at Loterios in Italy, prior to cleaning and etching. The ICV is suspended from the top dome of the OCV. 1-ICV; 2-middle section OCV; 3-top dome section OCV; 4-ICV weir drain port; 5-OCV cathode high voltage port.} 
\label{fig:CryostatAssembly} 
\end{figure}

The Xenon Detector and its LXe payload are 
contained in the Inner Cryostat Vessel. The Outer Cryostat Vessel 
provides its vacuum jacket. As shown in Fig.~\ref{fig:CryostatAssembly},
the OCV is supported at the bottom by three legs. The same assembly
provides shelves for the GdLS AVs
located underneath the OCV. The ICV is suspended from the top head of
the OCV with a mechanism enabling its levelling from above. 
Three long tubes run vertically to deploy calibration sources
into the insulating vacuum space between the vessels (see Sec.~\ref{sec:Calibrations}). 
Both vessels were designed in 
compliance with the ASME Boiler and 
Pressure Vessel Code Section VIII Div. 1. 

The ICV consists of a top head and a bottom vessel connected by a 
large flange near the top. The maximum outer 
diameter of the ICV is constrained 
by the cross-section of the Yates
shaft. Its tapered shape is to reduce the electric field near the cathode.
The TPC structure is anchored to the bottom of the ICV 
through six dedicated ports in
the dished end. Three angled ports below the main flange are provided
for the LXe weir drain return lines. 
Two ports at the top head and the central port
at the bottom are for the PMT and instrumentation cables.  The high voltage 
port has been machined on the inside to form a curvature minimizing the 
electric field around the cathode HV feed-through. Five plastic blocks are 
attached  to the tapered part of the ICV wall to prevent the ICV from
swinging  during a seismic event.     

The OCV consists of three segments in order to fit the 
largest into the conveyance of the Yates shaft. A port in the 
center of the top head hosts the low energy neutron source 
deployed in a large tungsten ``pig" (see Sec.~\ref{sec:Calibrations}). 
A reinforcing ring allows the top 
AVs to rest on the OCV head.   

The entire cryostat assembly is made out of a carefully selected
ultra-radiopure Grade-1 titanium sourced from a single titanium 
supplier~\cite{Akerib:2017iwt}. 
After a comprehensive material search campaign, a 5 metric ton 
Titanium Gr-1 slab was procured from Timet and used to 
fabricate all the stock material 
required for the cryostat. Initially it was cut into 
three pieces in order to roll the plates with multiple 
thicknesses, forge all the flanges, and the ports and 
to draw the welding wires. The ICV and OCV were 
fabricated from this material at Loterios in Milan, Italy (see
Fig.~\ref{fig:CryostatAssembly}). The cleaning and etching 
of the ICV and OCV is described in Sec.~\ref{sec:Materials}.

The ICV is maintained at its operating temperature by a set of 
closed-loop thermosyphon heat pipes
utilizing nitrogen as the process fluid. The thermosyphons deliver heat
from the ICV 
to the Cryogen On Wheels (COW), a bath of LN located 
above the water tank in the Davis Cavern. A cryocooler, model
SPC-1 from DH industries, re-condenses the boil-off nitrogen from 
the COW and transfers the heat to the chilled
water system. 
During cryocooler maintenance and repair,
the COW can be filled by transporting LN to the Davis Cavern
from the surface. Four 450 liter
storage dewars located underground,
act as an intermediate LN repository to enable this mode of operation.

Six copper coldheads are bolted to welded titanium fins on the ICV 
exterior and are serviced by three thermosyphon lines. 
The coldheads are placed
at a height just below the LXe level.
The cooling power of each
thermosyphon is set by adjusting the amount of process
nitrogen in each circuit. Fine adjustment is provided by PID-controlled
trim heaters located on each coldhead. Two additional
thermosyphon circuits remove heat from the LXe tower 
(see Sec.~\ref{sec:XeHandling}).

The total heat budget of the experiment is
estimated to be 700~W. The largest contributing item,
at 349~W, is due to the inefficiency of the primary two-phase xenon 
circulation heat exchanger. 
The thermosyphon trim heaters and the heat leak into the
ICV each account for about 115~W.

\subsection{Online Xe handling and purification}
\label{sec:XeHandling}

The online Xe purification system  
continuously removes electronegative impurities from the Xe while also
providing some measure of Rn removal and control.
Rejection of electronegatives begins during the final assembly of
the detector with a TPC outgassing campaign 
described in Sec.~\ref{sec:UGAssembly}. The electron 
lifetime goal is 800 $\mu$s, sufficient to drift charge 
from the cathode to the anode while suffering an acceptable
signal reduction factor of 1/e.

\begin{figure*}[t!]
\centering
\includegraphics[trim={0 0.0cm 0 0.0cm},clip,width=0.75\linewidth]{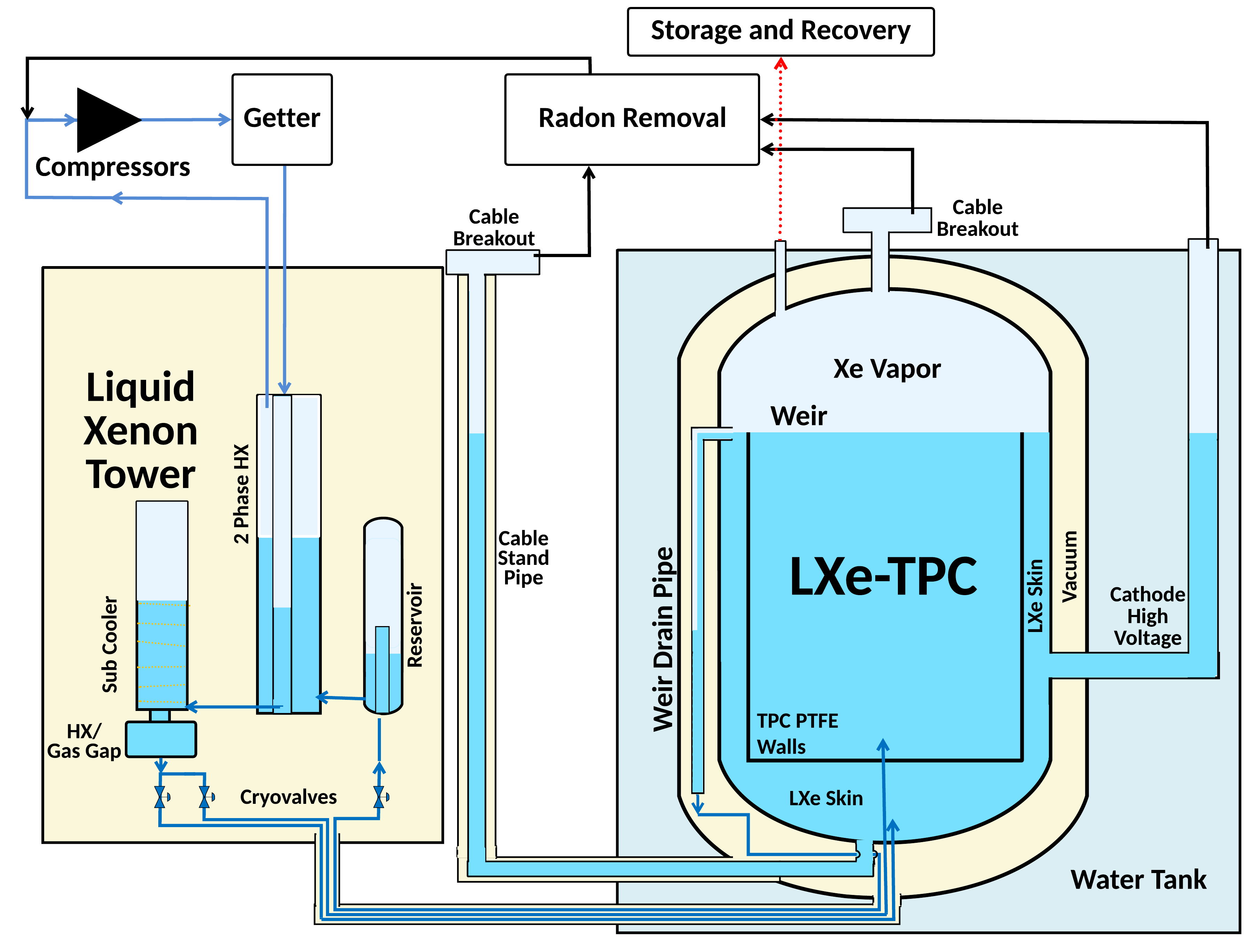}
\caption{Overview of the online Xe purification system. LXe
in the Xenon Detector (right) spills over a weir drain and 
flows horizontally to the Liquid Xenon tower, which stands outside
the water tank. It is vaporized in a two phase heat exchanger, 
pumped through a hot zirconium getter, and returned to the detector
after condensing. Cryovalves control the flow of LXe between 
the LXe tower and the Xenon Detector. A radon removal system 
treats Xe gas in the cable conduits and breakout feed-throughs
before sending it to the compressor inlet.}
\label{fig:XeCirculation} 
\end{figure*}

An overview of the system
is shown in Fig.~\ref{fig:XeCirculation}. Xe gas
is pumped through a hot zirconium getter at a 
design flow rate of 500 standard liters per 
minute (SLPM), taking 2.4~days
to purify the full 10 tonne Xe inventory in a single pass. 
The getter, model PS5-MGT50-R from SAES~\cite{SAES},  
operates at 400 $^{\circ}$C. For thermal efficiency, the getter
features a heat exchanger to couple the inlet and outlet 
Xe gas streams, substantially reducing the 3~kW heat burden at 
500 SLPM. A pre-heater ensures that the gas strikes
the getter bed at the operating temperature. Besides electronegative
removal, the getter bed also serves as a permanent repository 
for the tritium and $^{14}$C radio-labeled methane species 
that calibrate the beta decay response of 
the TPC (see Sec.~\ref{sec:Calibrations}).

Circulation flow is established by two all-metal 
diaphragm gas compressors,
model A2-5/15 from Fluitron~\cite{fluitron}. %See Fig.~\ref{fig:XeCompressor}. 
The two compressors operate in parallel, each
capable of 300 SLPM at 16 PSIA inlet pressure.
The system operates
with one compressor at reduced flow rate during periodic maintenance.
The total achieved gas flow is trimmed by a bellows-sealed
bypass proportional valve, model 808 from RCV.
Both circulation compressors have two stages, 
each featuring copper seals plated onto stainless steel 
diaphragms. All-metal sealing technology was chosen to 
limit radon ingress from air.

The LXe tower is a cryogenic device standing 
on the floor of the Davis Cavern 
outside the water tank and at a height somewhat
below the Xenon Detector. Its primary purpose 
is to interface the liquid and gaseous portions of the online
purification circuit and to efficiently exchange heat between them. There are four vessels in the tower:
the reservoir vessel, the two-phase heat exchanger (HEX), 
the subcooler vessel, and the subcooler HEX. 

The reservoir vessel collects LXe departing 
the Xenon Detector via the weir system. It features a standpipe 
construction to decouple its liquid level from that 
in the weir drain line. 
LXe flows from the bottom of the reservoir into the 
two-phase HEX, where it vaporizes after exchanging 
heat with purified Xe gas returning from the getter.
The two-phase HEX is an ASME-rated brazed plate 
device made by Standard Xchange 
consisting of corrugated stainless steel.
On its other side, condensing LXe flows into 
the subcooler vessel and subcooler HEX. The vessel
separates any remaining Xe gas from the LXe, while
the HEX cools the LXe to below its saturation temperature.
The HEX consists of three isolated elements: 
the LXe volume, an LN thermosyphon coldhead 
cooled to 77~K, and a thin thermal coupling gap. 
The power delivered to the LXe 
can be varied from 90 W and 480 W by adjusting the composition 
of the He/N$_2$ gas mixture in the gap.
An additional thermosyphon coldhead integrated with the 
reservoir removes excess heat
during cooldown and operations.
Both the reservoir and the subcooler vessels 
are equipped with LXe purity monitors (LPMs) to monitor
electronegatives entering and exiting the Xenon Detector.  
Each LPM is a small, single-phase TPC which drifts free electrons 
over a distance, and measures the attenuation of the electrons 
during that transit. 

LXe flows between the LXe tower and the Xenon Detector through 
three vacuum insulated transfer lines that run across the bottom
of the Davis Cavern water tank. Two lines connect to the bottom
of the ICV and supply sub-cooled LXe to the TPC and 
skin regions of the Xenon Detector. The third line 
returns LXe from the ICV weir drain 
system to the reservoir. The lines are constructed
by Technifab with an integrated vacuum insulation jacket. 
They are further insulated 
from the water by an additional vacuum shell.
Cryogenic control valves 
from WEKA regulate the LXe flow in each of the three lines.

Conduits connect to the ICV at its 
lower flange and upper dome to service PMT and instrumentation cables
to the TPC. The lower conduit, which is vacuum insulated and filled with 
LXe, travels across the water tank floor, 
penetrates its side wall, and mates with a 
vertical LXe standpipe. Its cables emerge 
into gaseous Xe and then connect to breakout feed-throughs
at the standpipe top. Two upper conduits filled with gaseous
Xe connect the ICV top head 
to breakout feed-throughs and service cables
to the upper part of the Xenon Detector.

The Xe gas in the cable conduits and breakout feed-throughs are treated
for radon removal by a cold synthetic charcoal column 
drawing 0.5 SLPM of Xe gas flow. The system is designed to 
sequester $^{222}$Rn for three half-lives, 
or 12.7 days, allowing 90\% of these atoms to decay. 
The sequestration is accomplished by a gas chromatographic process that 
employs 10 kilograms of synthetic charcoal (Saratech Spherical Adsorbent, 
Blücher GmbH) cooled to 190~K~\cite{Pushkin:2018wdl}. 
The technique was previously 
demonstrated in Ref.\cite{abe201250}. 
The charcoal was etched in nitric acid and rinsed with distilled water
to reduce its radon emanation rate. 

Besides the LPMs, surveillance of the 
impurity content of the Xe is also provided
by two coldtrap mass spectrometry 
systems~\cite{leonard:2010zt,dobi:2011vc}. 
These devices monitor for the presence 
of stable noble gas species such as
$^{84}$Kr and $^{40}$Ar and also for electronegatives
such as O$_2$.
Ten standard liter samples of Xe gas are collected 
and passed through a coldtrap cooled to 77~K, 
a temperature at which Xe is retained while many impurities species
pass through. The outlet of the coldtrap is monitored
by a Residual Gas Analyzer (an RGA200 from SRS). 
The sensitivity for detecting $^{84}$Kr in Xe
is better than 10 parts-per-quadrillion (g/g).  
The coldtrap is cooled either with a pulse tube
refrigerator (model PT60 from Cryomech Inc.) or with an
open flask dewar of liquid nitrogen. One of these systems is 
permanently plumbed to fixed locations
in the Xe handling system; the other acts as a mobile 
utility system to be deployed as needed. Both are highly automated
and allow for multiple measurements per day.

To recover the ten tonne Xe 
inventory to long term storage, 
two high pressure gas compressors 
(Fluitron model D1-20/120) pump Xe gas into 12 Xe 
storage packs. The recovery compressors use the same all-metal 
diaphragm technology as the circulation compressors. 
Each Xe storage pack consist of 12 
DOT-3AA-2400 49.1 liter cylinders sealed with 
Ceoduex D304 UHP tied diaphragm valves and ganged 
together in a steel frame. Each pack 
weighs 1,800~kg when full. During Xe recovery 
heat is added to the LXe 
by electrical heaters and by softening the insulating 
vacuum of the lower cable conduit. Two backup diesel 
generators are provided in case of a power outage. 
The emergency recovery logic is described in Sec.~\ref{subsec:Controls}.

All elements of the online Xe handling system were cleaned
for ultra high vacuum with solvents and rinsed in de-ionized water. 
Orbital welds conform to ASME B31.3. Where possible, 
stainless steel components have been etched in citric acid
to reduce radon emanation.

\subsection{Removal of Kr from Xe}

Beta decay of $^{85}$Kr in the LXe is a challenging
ER background source.
The acceptable Kr concentration, derived by 
assuming an isotopic abundance of
$^{85}$Kr/Kr $\sim 2\times10^{-11}$, is 
Kr/Xe $<$ 0.3~parts-per-trillion (g/g).
This concentration is achieved prior to the start of LZ operations 
by separating trace Kr from the Xe inventory
with a gas charcoal chromatography process. 

A total of 800~kg of Calgon PCB activated 
charcoal is employed, divided evenly into two 
columns. The charcoal was washed with water 
to remove dust and baked under an N$_2$ purge
for 10 days, ultimately achieving 
a charcoal temperature of 150 $^\circ$C.
During processing the Xe inventory is mixed with He 
carrier gas circulated by a compressor 
(model 4VX2BG-131 from RIX). 
The Xe/He mixture is passed
through one of the two charcoal columns 
at a pressure of 1.7 bara. 
Trace Kr reaches the column outlet first and 
is directed to an LN-cooled charcoal trap where it 
is retained. 
A Leybold-Oerlikon Roots blower and screw pump located 
at the charcoal column outlet then activates, dropping 
the column pressure to 10~mbar. This purges
the purified Xe from the column. The Xe is
separated from the He carrier gas by freezing it 
at 77~K in an LN-cooled heat exchange vessel. 
This freezer is periodically warmed to vaporize
the Xe ice, and the recovered
Xe gas is transferred at room temperature by 
a Fluitron Xe recovery compressor 
to one of the twelve Xe storage packs described above.

The entire 10 tonne 
Xe inventory is processed in 16~kg batches. The chromatographic
and Xe purge cycles each take about 2 hours.
Two charcoal columns are employed to allow processing
and column purging to proceed in parallel. 
A Kr rejection factor greater than
1000 can be achieved in a single pass through the system; 
two passes are envisioned to achieve the required concentration.
The processing is monitored by a coldtrap Xe mass spectrometry 
system for quality assurance. After processing, the Xe storage
packs are shipped from SLAC to SURF in preparation for 
condensing into the Xenon Detector.

\section{Outer Detector}
\label{sec:OD}
The principal purpose of the Outer Detector is to tag neutrons scattering events in the TPC. Most neutrons originate from radioactive impurities in material immediately adjacent to the TPC, such as those from ($\alpha$,n) processes on PTFE. The OD is a near-hermetic liquid scintillator detector designed to capture and tag neutrons within a time window that allows the signals to be correlated with the NR in the TPC.

The detection medium for the OD is gadolinium-doped liquid scintillator contained within segmented acrylic vessels that surround the OCV.
Neutrons are detected predominantly through capture on $^{155}$Gd and $^{157}$Gd; a total of 7.9 MeV ($^{155}$Gd) or 8.5 MeV ($^{157}$Gd) is released in a post-capture cascade of, on average, 4.7 gammas. About 10\% of neutrons capture on hydrogen, emitting a single 2.2~MeV gamma. Gammas induce scintillation within the LS which is subsequently collected by the 120 8--inch PMTs that view the OD from a support system inside the water tank. To maximize light collection efficiency, there is both a Tyvek curtain behind, above and below the PMTs, and a layer of Tyvek surrounding the cryostat. 

\begin{figure}[h!]
\centering
\includegraphics[trim={0 0.0cm 0 0.0cm},clip,width=0.95\linewidth]{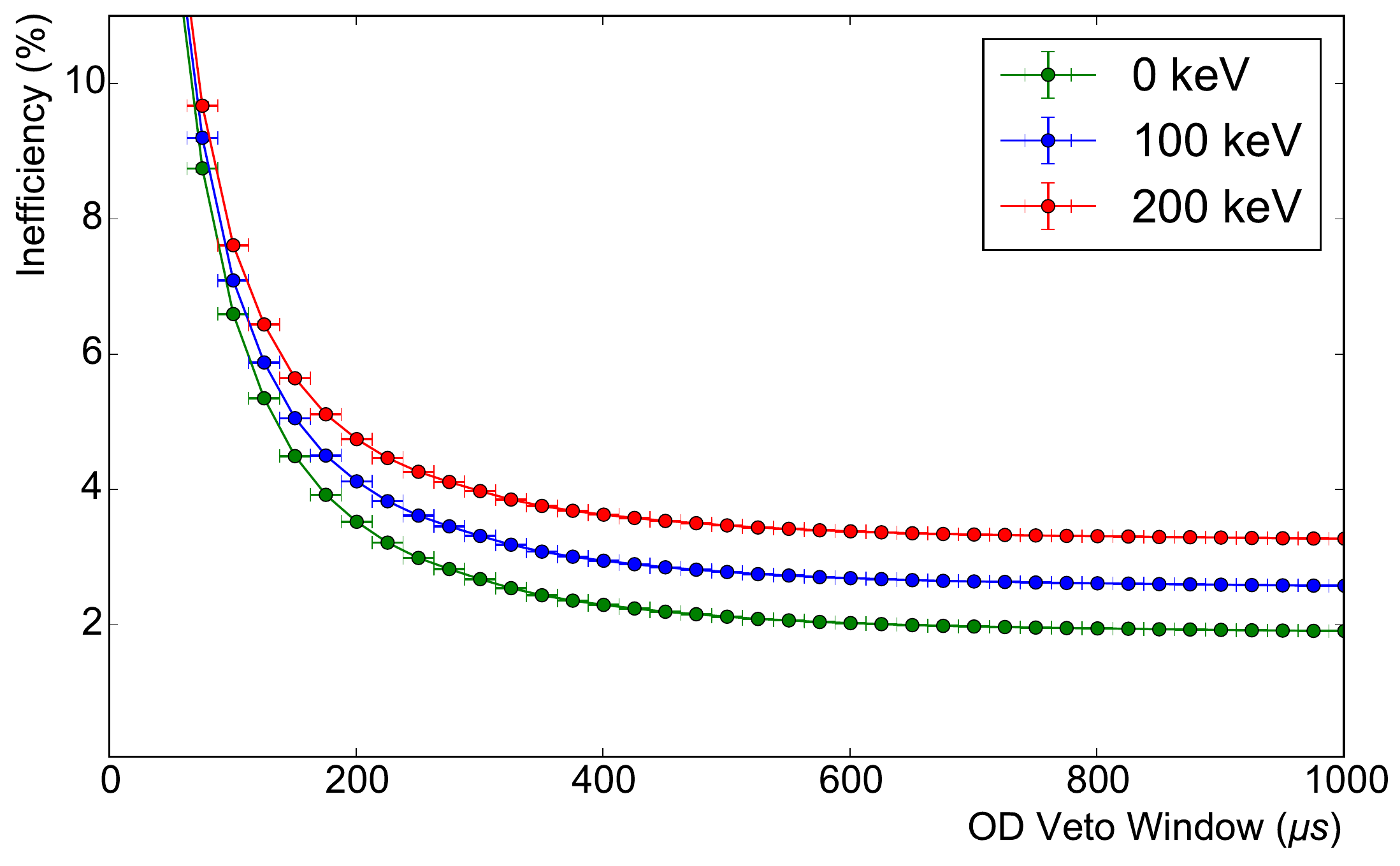}
\caption{Monte Carlo derived OD inefficiency as a function of veto window (time between S1 in the TPC and signal in the OD). The energy thresholds referenced in the legend are for electron recoils.  \label{fig:ODInefficiency}}
\end{figure}

The OD has been designed to operate with a neutron detection efficiency of greater than 95\%. To optimize this efficiency, the concentration of Gd was chosen such that capture on H is sub-dominant. Furthermore, the time between a signal in the TPC and a neutron capture in the OD impacts the efficiency (see Fig.~\ref{fig:ODInefficiency}). The level of Gd chosen for LZ reduces the average capture time of thermal neutrons in liquid scintillator from 200~$\mu$s to 30~$\mu$s. However, there is a significant population of neutrons that survive several times longer than 30~$\mu$s. Simulations demonstrate that neutrons can spend significant time scattering and thermalizing within the acrylic walls of the OD vessels.  To minimize
this effect, the acrylic walls are designed to be as thin as is structurally possible. Using less acrylic also reduces the number of H-captures. The use of a 500~$\mu$s time window allows for an efficiency of 96.5\% for a 200 keV threshold, while achieving a deadtime of less than 5\%. 

\subsection{Outer Detector systems}

\begin{figure}[t!]
\centering
\includegraphics[trim={0 0.0cm 0 0.0cm},clip,width=0.85\linewidth]{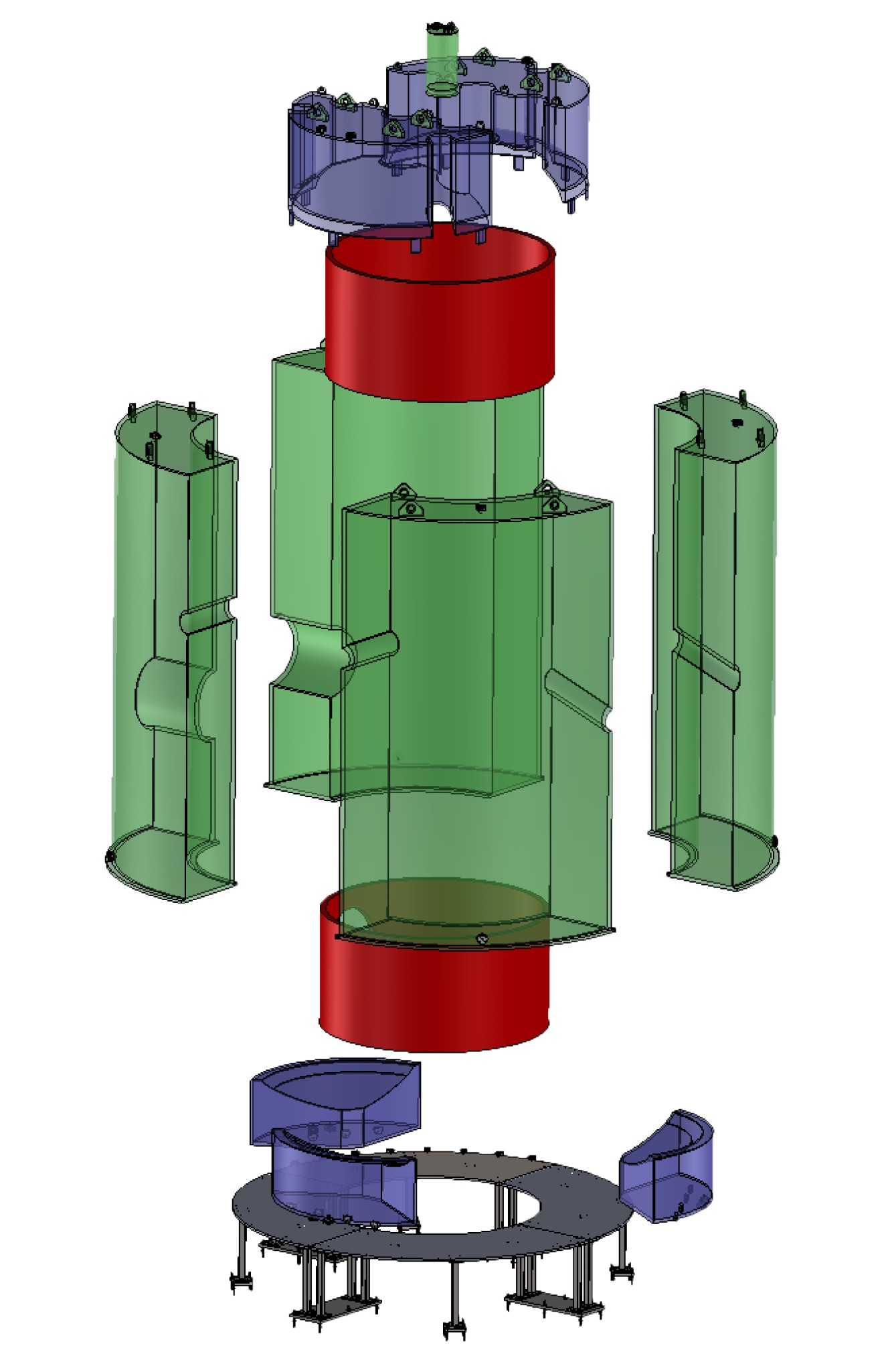}
\caption{The outer detector system in an exploded view.  The four large side vessels are shown in green and the 5 smaller top and bottom vessels are shown in blue.  Also shown are water displacers in red, and the stainless steel base in grey.}
\label{fig:explodedOD} 
\end{figure}
A total of ten ultra-violet transmitting (UVT) Acrylic Vessels have been fabricated by Reynolds Polymer Technology in Grand Junction, Colorado. These consist of four side vessels, three bottom vessels, two top vessels and a `plug` which can be removed for photoneutron source deployment, see Fig.~\ref{fig:explodedOD}. The AVs were designed as segmented to allow transport underground and into the water tank with no acrylic bonding necessary on site. 
All acrylic walls for the side AVs are nominally 1--inch thick. For the top and bottom AVs, the side and domed acrylic walls are 0.5--inch thick, whereas the flat tops and bottoms are 1--inch thick for structural reasons. The AVs contain various penetrations for conduits and cabling. 
All sheets of acrylic used for fabrication were tested for optical transmission and were found to exceed 92\% between 400 and 700~nm, meeting requirements. Acrylic samples were screened with ICP-MS (see Sec.~\ref{sec:HPGeMS})and found to be sufficiently low in radioactive contaminants. 

The liquid scintillator used in the OD consists of a linear alkyl benzene (LAB) solvent doped at 0.1\% by mass with gadolinium. The full chemical make-up of the GdLS cocktail is shown in Table~\ref{tab:GdLS}. 
Gd is introduced into LAB using trans-3-Methyl-2-hexenoic acid (TMHA) as a chelation agent. Other components are the fluor, 2,5-Diphenyloxazole (PPO), and a wavelength shifter, 1,4-Bis(2-methylstyryl)-benzene (bis-MSB). The emission spectrum of this mix spans 350 to 540~nm, with peaks at 400 and 420~nm, and the absorption length in this range is of order 10~m. 
LAB, TMHA and PPO are purified in order to remove metallic and coloured impurities to improve optical transmission; LAB and TMHA by  thin-film distillation and PPO by water-extraction and recrystallization. For the chelated Gd product, a self-scavenging method is used to induce precipitation of uranium and thorium isotopes to improve radiopurity.
Twenty-two tonnes of GdLS contained in 150 55-gallon drums are shipped to SURF and transferred into the AVs through a reservoir. Exposure to air is minimized, as oxygen, radon and krypton negatively impact the GdLS performance. The GdLS is bubbled with nitrogen while in the reservoir in order to remove dissolved oxygen and maximize the light yield. Furthermore, a light yield test is performed on each drum of GdLS before transfer into the AVs. The test apparatus consists of a dark box containing a radioactive source, one PMT, and a small sample of the GdLS.

\begin{table*} [ht]
\caption{Chemical components in \SI{1}{\liter} of GdLS.}
\centering
\begin{tabular} 
{c c c c} 
\hline
    {\bfseries Acronym} &
    {\bfseries Molecular Formula} &
    {\bfseries Molecular Weight (g/mol)} & 
    {\bfseries Mass (g)} \\
\hline
\vphantom{\Large L}LAB & C$_{17.14}$H$_{28.28}$ & 234.4 & 853.55 \\
PPO & C$_{15}$H$_{11}$NO& 221.3 & 3.00 \\
bis-MSB & C$_{24}$H$_{22}$ & 310.4 & 0.015 \\
TMHA & C$_9$H$_{17}$O$_2$ & 157.2 & 2.58 \\ 
Gd & Gd & 157.3 & 0.86 \\ \hline
\vphantom{\Large G}
GdLS & C$_{17.072}$H$_{28.128}$O$_{0.0126}$N$_{0.0037}$Gd$_{0.0015}$ & 233.9 & 860.0 \\ \hline
\end{tabular}
\label{tab:GdLS}
\end{table*}

A total of 120 Hamamatsu R5912 8--inch PMTs view the GdLS and AVs from a Tyvek curtain situated 115~cm radially from the outer wall of the acrylic (see Fig.~\ref{fig:LZSolid}). The interaction rate in the OD from radioistopes in the PMT system is predicted to be only 2.5~Hz due to the shielding provided by the water gap. The PMTs are arranged in 20 ladders spaced equally around the water tank in $\phi$ with 6 PMTs per ladder. Each PMT is held by a `spider' support and covered by a Tyvek cone.

A dedicated Optical Calibration System (OCS) has been designed for PMT monitoring and measurement of the optical properties of the GdLS and acrylic. Thirty LED duplex optical fibres are mounted with the PMT supports, with an additional five beneath the bottom AVs placed specifically to check transmission through the acrylic. Short, daily calibrations with the OCS will be performed in order to check the PE/keV yield at the veto threshold, and weekly calibrations will be used to check PMT gains and optical properties. 

\subsection{Performance}
The performance of the OD strongly depends on its event rate. The sources can be divided into internal and external; internal backgrounds are contamination intrinsic to the OD, i.e. inside the GdLS; while external sources can be subdivided again into radioactivity from LZ components and radioactivity from the Davis Cavern itself (see Table~\ref{tab:ODrate}).  

Radioactive contaminants internal to the GdLS have been measured through a campaign with an LS Screener, a small detector containing 23~kg of liquid scintillator viewed  by three low background LZ PMTs,  fully described in Ref.~\cite{Haselschwardt:2018}. The LS Screener took data with both loaded (with Gd) and unloaded (no Gd) samples of the liquid scintillator that is used in the Outer Detector. The unloaded sample allowed a clear determination of what contaminants were introduced during the Gd-doping process, as well as a clearer low energy region to allow a measurement of the $^{14}$C concentration, particularly important as its rate influences the choice of energy threshold. Use of pulse shape discrimination allowed for efficient separation of alpha decays from betas and gammas, and constraints were placed on activity from the $^{238}$U, $^{235}$U and $^{232}$Th decay-chains, $^{40}$K, $^{14}$C, $^{7}$Be (from cosmogenic activation of carbon in the LS), $^{85}$Kr and $^{176}$Lu. Some surprising and significant findings of the LS Screener were the dominance of the rate from nuclides within the $^{235}$U chain, and the presence of $^{176}$Lu, now known to be introduced when doping with gadolinium, since neither were observed in the unloaded LS sample. A more aggressive purification of the GdLS resulted in a decrease in activity of almost all contaminants. The new, lower activities were used in combination with the LS Screener results to predict a rate of 5.9 Hz above the nominal 200~keV veto threshold for the OD. 

\begin{table}[h]
\centering
\footnotesize
\caption{\small Predictions for the event rate in the Outer Detector. Rates are given in the case of the nominal 200~keV threshold. \label{tab:ODrate}}
\begin{tabular}{ l l c } \hline 
\textbf{Type} & \textbf{Component} & \textbf{OD Rate (Hz)} \\  \hline 
\multirow{6}{*}{External} & PMTs \& Bases & 0.9 \\
& TPC & 0.5 \\
& Cryostat & 2.5 \\
& OD & 8.0 \\
& Davis Cavern & 31 \\    \hline 
Internal & GdLS & 5.9 \\   \hline   
 \textbf{Total} & & \textbf{51} \\ \hline  
 
\end{tabular}
\end{table} 

The biggest contribution to the rate in the OD is from the radioactivity within the Davis Cavern. Contamination of the cavern walls with on the order of tens of Bq/kg for $^{40}$K, $^{238}$U and $^{232}$Th has been established using dedicated measurements of the $\gamma$-ray flux with a NaI detector~\cite{Akerib:2019sek}, and simulation studies suggest a rate above 200 keV of $27\pm7$~Hz, concentrated in the top and bottom AVs.  

With an expected overall rate of $\sim$50~Hz, the OD can be expected to operate with an efficiency of 96.5\% for a 200~keV threshold. The energy threshold of the OD is nominally a number of photoelectrons corresponding to an energy deposit of 200~keV, predicted to be 10~PE by photon transport simulations. The threshold is chosen to eliminate the rate from internal $^{14}$C contamination, as it is a low energy $\beta$-decay with an endpoint of 156~keV. The OD may be operated instead with a 100~keV threshold, depending on the observed rate, which would decrease the inefficiency at a window of 500~$\mu$s from 3.5\% to 2.8\%. 

The impact of the OD on NR backgrounds is characterized through neutron Monte Carlo simulations. The total NR background in 1000 livedays is predicted to be reduced from 12.31 to 1.24 NR counts when the OD and skin vetoes are applied, with the OD providing most of the vetoing power. Due to the spatial distribution of these NRs in the LXe TPC, the OD is necessary to utilize the full 5.6 tonne fiducial volume.

\section{Calibrations}
\label{sec:Calibrations}

\newcommand{\isot}[2]{$^{\textrm{#2}}$#1}

Many attributes of the LZ detector response require \emph{in situ} calibration.  Calibration goals range from low-level quantities such as PMT gain and relative timing to high-level quantities such as models of electron recoil and nuclear recoil response.  To these ends, the LZ detector includes significant accommodations for a suite of calibrations.  Large-scale accommodations (some visible in Fig.~\ref{fig:LZSolid}) include three small-diameter conduits to transport external sources to the cryostat side vaccum region, one large-diameter conduit to transport large ($\gamma$,n) sources to the cryostat top, and two evacuated conduits to enable neutron propagation from an external neutron generator to the cryostat side.

\renewcommand{\arraystretch}{1.1}
\begin{table} [t]
\caption{Overview of radioactive nuclide sources planned for LZ calibration, grouped according to deployment method.  A: gaseous sources released into GXe circulation, B: sealed sources lowered down small-diameter conduits to cryostat side vacuum, C: ($\gamma$,n) sources requiring dense shielding, lowered down a large-diameter conduit to the cryostat top, D: DD generator sources, in which neutrons travel through conduits from the generator, through the water tank and outer detector.}

\centering
\begin{tabular} 
{|l|l|l|l|l|} 
\hline
& Nuclide & Type & Energy [keV] & $\tau_{1/2}$ \\
\hline
& \isot{Kr}{83m} & $\gamma$ & 32.1 , 9.4 &1.83~h  \\ 
& \isot{Xe}{131m} & $\gamma$ & {164}  &11.8~d\\
A & $^{220}$Rn & $\alpha, \beta, \gamma$ & various & 10.6~h  \\
& $^3$H & $\beta$ &  18.6 endpoint &12.5~y \\
& $^{14}$C & $\beta$ & 156 endpoint &5730~y\\
\hline
& $^{241}$AmLi & ($\alpha$,n) & 1500 endpoint $^{(a)}$ &432~y   \\ 
& $^{252}$Cf & n & Watt spectrum &2.65~y \\
& $^{241}$AmBe & ($\alpha$,n) & 11,000 endpoint &432~y \\
&$^{57}$Co & $\gamma$ & {122} &0.74~y  \\
B &$^{228}$Th & $\gamma$ & {2615} &1.91~y  \\
&$^{22}$Na & $\gamma$ & {511,1275} &2.61~y   \\
&$^{60}$Co & $\gamma$ & 1173 , 1333 &5.27~y \\
&$^{133}$Ba & $\gamma$ & {356} &10.5~y  \\
&$^{54}$Mn & $\gamma$ & {835}  &312~d  \\
\hline
& $^{88}$YBe & ($\gamma$,n) & 152 &107~d  \\ 
C & $^{124}$SbBe & ($\gamma$,n) & 22.5 &60.2~d \\
& $^{205}$BiBe & ($\gamma$,n) & 88.5 &15.3~d    \\
& $^{206}$BiBe & ($\gamma$,n) & 47 &6.24~d   \\
\hline
D & DD &  n & 2450  &$-$ \\
& D Ref. & n & $272\rightarrow400$ &$-$ \\
\hline

\end{tabular}
\label{table:sourcelist}
\end{table}

\subsection{Internal sources}

Gaseous sources can mix with the LXe in order to reach the central active volume, where self-shielding limits calibration via external gamma sources.  The baseline suite of such `internal' sources is listed in Group A of Table~\ref{table:sourcelist}.

Long-lived gaseous sources (\isot{H}{3}, \isot{C}{14}) can be stored as a pressurized gas, with purified Xe serving as the carrier.  Because the nuclide is long-lived, it must be in a chemical form that can be efficiently removed by the getter (see Sec.~\ref{sec:XeHandling}).  The LZ implementation builds on the successful example of LUX, in which isotopically-labeled CH$_4$ served as the calibration gas.  CH$_4$ was seen to be efficiently removed, as long as it did not contain trace amounts of other labeled hydrocarbons~\cite{dobi2014, akerib:2015wdi}.

\begin{figure}[t!]
\centering
\includegraphics[trim={0 0.0cm 0 0.0cm},clip,width=0.80\linewidth]{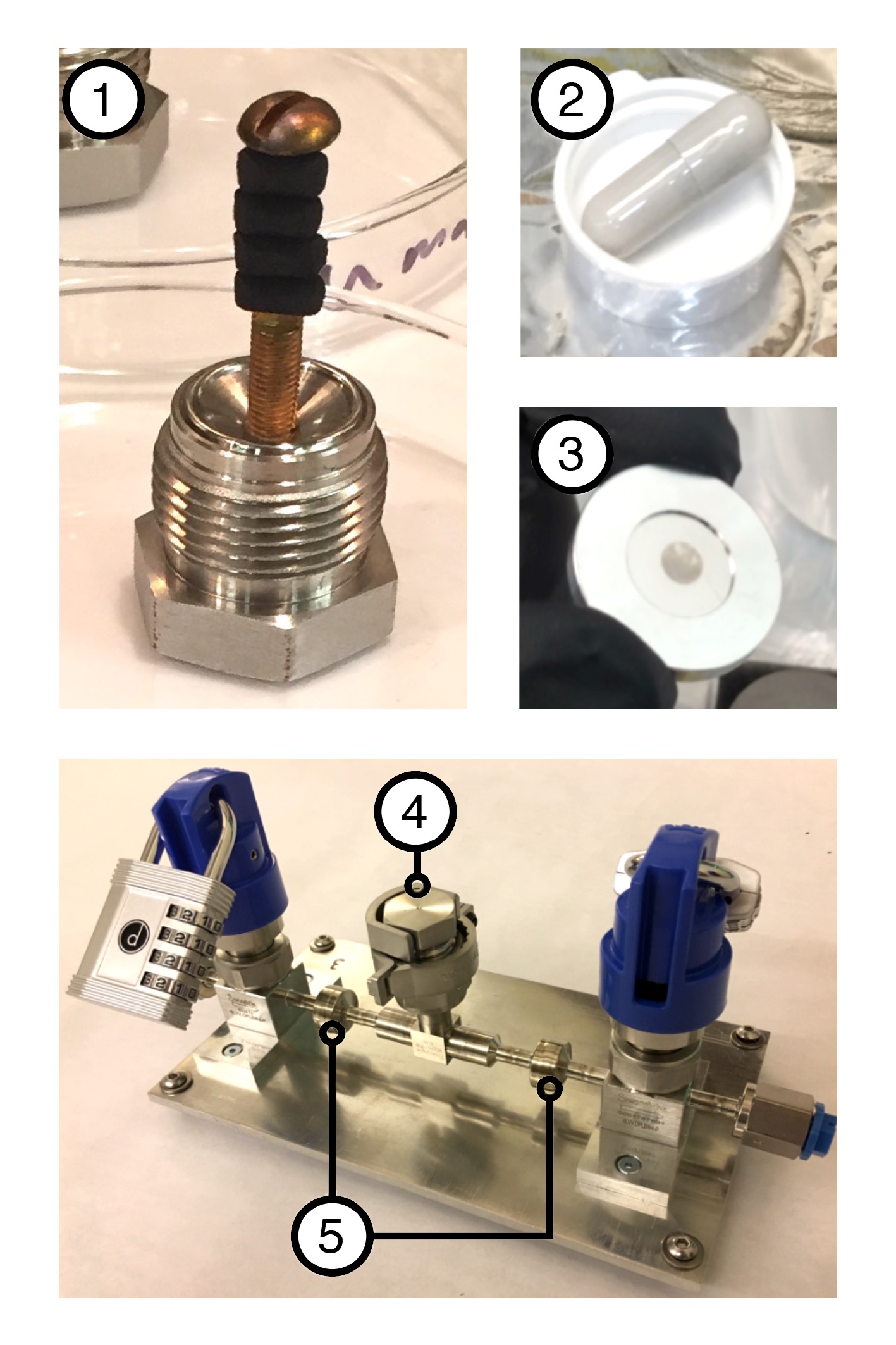}
\caption{TOP: Three example solid materials containing parent nuclides that emit daughter calibration gaseous sources.  Charcoal dosed with \isot{Rb}{83} is fixed to the a 1/2-inch VCR plug (1).  A gas-permeable pill (2) containing \isot{I}{131} and a disk source (3) of electroplated \isot{Th}{228} can also be fixed in place.  BOTTOM:  Photograph of a typical gaseous source generator.  Carrier Xe gas flows from left to right.  The active parent material is stored in the central region (4), accessed via a 1/2-inch VCR port.  This region is bounded by a pair of filter elements (5) of 3~nm pore size sintered nickel and then a pair of lockable manual valves for isolation during shipping and installation.}
\label{fig:generatorplumbing} 
\end{figure}

The short-lived gaseous sources (\isot{Kr}{83m}, \isot{Xe}{131m}, $^{220}$Rn) are stored in the form of their parent nuclide, which can be handled and stored in a compact solid form and placed within `generator' plumbing in which it emanates the calibration daughter.  \isot{Rb}{83} serves as the parent nuclide of \isot{Kr}{83m}, and is deposited in aqueous solution on high purity, high surface area charcoal before baking (as in~\cite{akerib:2017eql}).  \isot{I}{131} serves as the parent nuclide of \isot{Xe}{131m}, and is commercially available in a pill form of high Xe emanation efficiency.  \isot{Th}{228} serves as the parent nuclide of \isot{Rn}{220}, and is available commercially from Eckert \& Ziegler as a thin electroplated film for optimal Rn emanation.  In the LZ implementation, these generator materials are housed in transportable and interchangeable plumbing sections (see Fig.~\ref{fig:generatorplumbing}).  These assemblies contain both a port for material access and a pair of sintered nickel filter elements (3~nm pore size, Entegris WG3NSMJJ2) to prevent contamination by the parent nuclide of the active Xe.

Both long-lived and short-lived gaseous sources require precise dose control on the injected activity, accomplished via a gas handling system dedicated to injection control.  A specific GXe cylinder supplies the carrier gas to transport small quantities of calibration gas through a series of high-precision Mass Flow Controllers (Teledyne-Hastings HFC-D-302B) and volumes of precise pressure measurement (MKS 872).  Once a dose of calibration gas has been isolated, the volume containing the dose is flushed into the main GXe circulation flow path, either before the getter for noble-element calibration species or after the getter for long-lived CH$_4$-based species.

\subsection{External sources}

\begin{figure}[h!]
\centering
\includegraphics[trim={0 0.0cm 0 0.0cm},clip,width=0.95\linewidth]{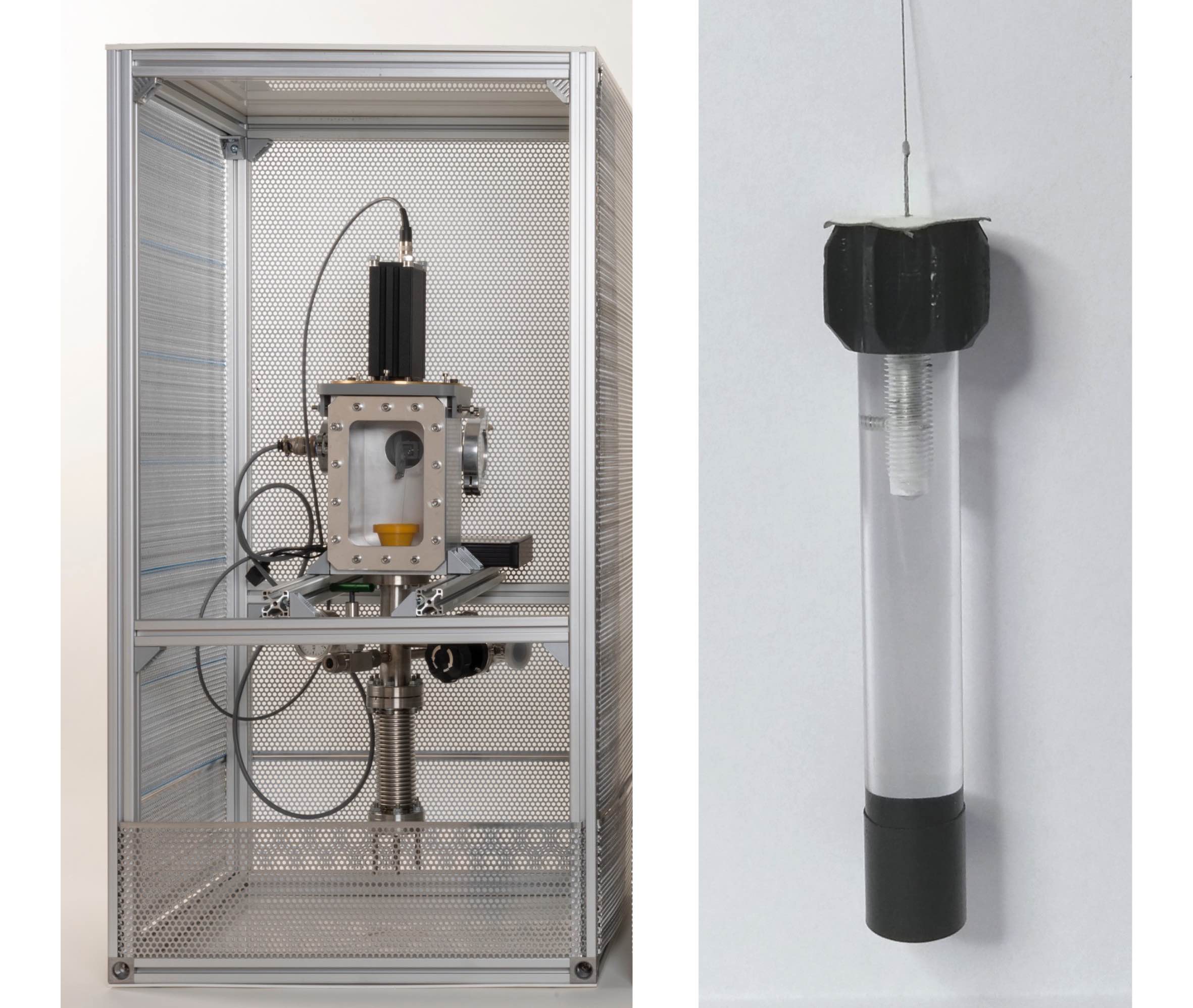}
\caption{LEFT: One of three external source deployment systems, including the laser ranging system (top black component) and the stepper motor and gear/winding assembly (enclosed in a KF50 Tee at the back).  A transparent plate makes visible the region in which the sources are installed and removed.  RIGHT:  An external source assembly, showing the acrylic body, the source region (at bottom), and the filament connection, black skids, and laser reflector (at top).}
\label{fig:sourcedeployment} 
\end{figure}

External sources are lowered through three 23.5~mm ID 6~m long conduits to the vacuum region between the ICV and OCV.  Each conduit is capped by a deployment system (Fig.~\ref{fig:sourcedeployment}) which raises and lowers the sources with final position accuracy of $\pm$5~mm.  The position measurement is accomplished via an ILR1181-30 Micro-Epsilon laser ranger (visible at the top of Fig.~\ref{fig:sourcedeployment}), supplying live data for an active feedback protocol to a SH2141-5511 (SANYO DENKO Co) stepper motor.  A $\sim$100~$\mu$m nylon composite filament suspends the sources, rated to a maximum load of 12~kg.  The external sources themselves are in most cases commercial sources (Eckert \& Ziegler type R).  A special case is the AmLi source, custom fabricated but of the same form factor.  To enable smooth transport up and down the conduit, each source is epoxied and encapsulated at the lower end of a 5" long by 0.625" diameter acrylic cylinder.  The top end contains the capsule holder allowing connection to the filament and includes a ferromagnetic connection rod for recovery in case of filament breakage.

\subsection{Photoneutron sources}

A selection of photoneutron ($\gamma$,n) sources, including $^{88}$YBe, $^{124}$SbBe, $^{204}$BiBe and $^{205}$BiBe, are planned to calibrate the nuclear recoil energy range from below 1 keV up to about 4.6 keV. This range corresponds to the expected energy depositions from \isot{B}{8} solar neutrino coherent scattering. Only about one neutron is produced for every $10^{4}$ gammas emitted, so a significant quantity of gamma shielding is required (see Fig.~\ref{fig:neutronsources}). The neutrons are quasi mono-energetic at production (within a few percent) but undergo additional scatterings before they reach the liquid xenon. The utility of this calibration source is derived from the endpoint energy the neutron deposits, which simulations indicate will be clearly distinguishable after a few days of calibration.

A ${\sim}$140 kg tungsten shield block is designed to be deployed at the top of LZ via a crane. In the unlikely event the shield block were to become lodged inside the LZ water tank, it would be possible to separately remove the conical structure which contains the gamma source and the Be.

\subsection{Deuterium-deuterium neutron sources}

\begin{figure}[h!]
\centering
\includegraphics[trim={0 0.0cm 0 0.0cm},clip,width=0.95\linewidth]{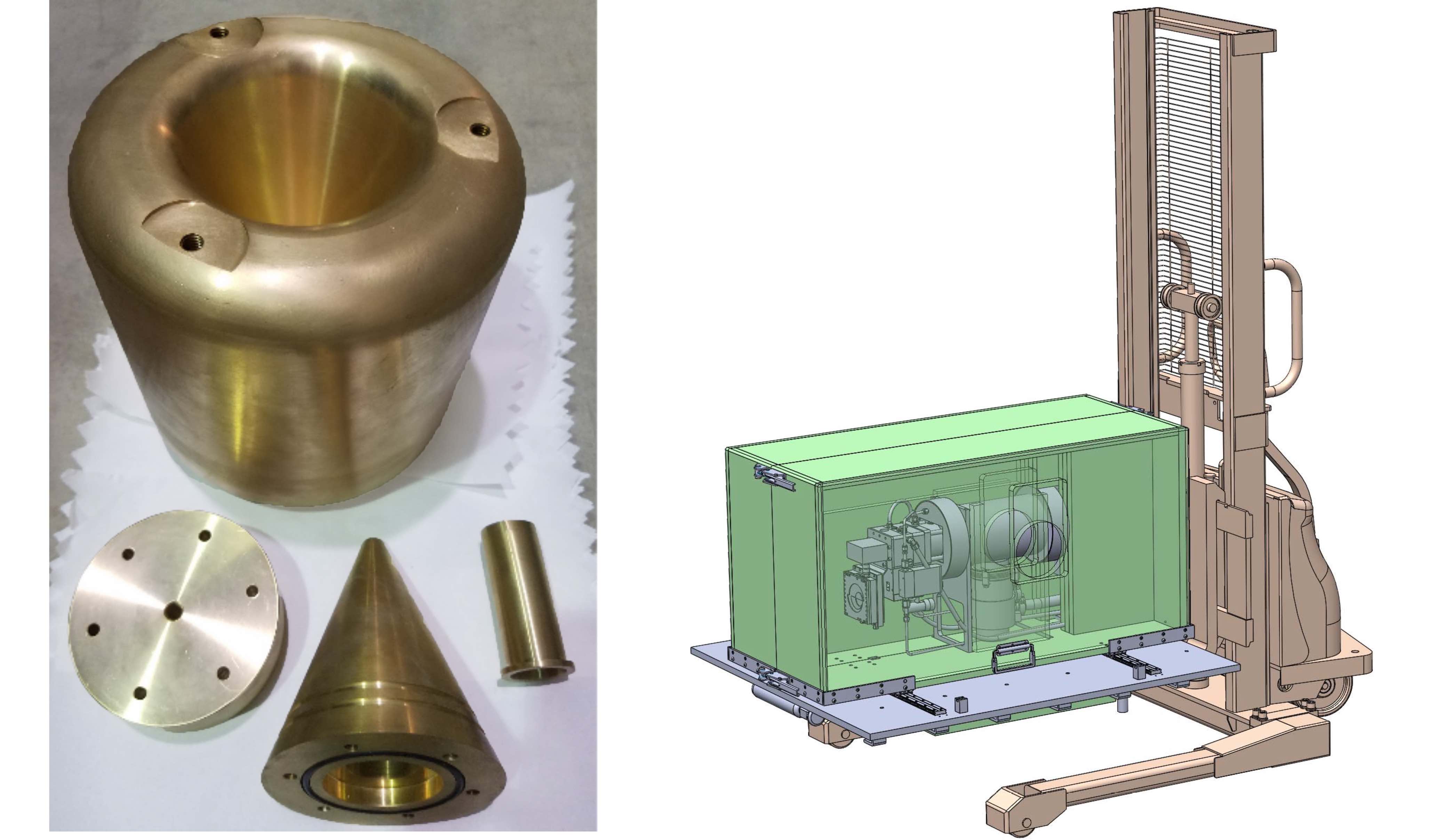}
\caption{LEFT:  Brass mockup of the photoneutron source shielding assembly, which will be made of tungsten alloy.  The main body is at top. Depicted below is a conical insert that houses radioactive source.  RIGHT:  Rendering of the DD generator in its boron-doped shielding assembly, all mounted on a movable positioning system.}
\label{fig:neutronsources} 
\end{figure}

An Adelphi DD-108 deuterium-deuterium (DD) neutron generator produces up to $10^8$ neutrons per second. A custom upgrade will allow up to $10^9$ n/s. The neutrons are delivered through the Davis Cavern water tank and Outer Detector via dedicated neutron conduits. There are two sets of conduits, one level and one inclined at 20 degrees from the horizontal. Each conduit assembly includes a 2--inch diameter and a 6--inch diameter path, and all are filled with water during dark matter search. As shown in Fig.~\ref{fig:neutronsources}, the generator is permanently mounted on an Ekko Lift (model EA15A), and surrounded by custom neutron shielding material. A kinematic mounting plate, located between the forks of the lift, will bolt to threaded inserts in the concrete floor. This is designed to provide precise, repeatable positioning.

The DD-108 produces 2450 keV mono-energetic neutrons. This source has already been used by LUX to obtain a precise, \emph{in-situ} calibration of the low-energy nuclear recoil 
response~\cite{akerib:2016mzi}. In addition to this mode of operation, LZ obtains 272~keV quasi mono-energetic neutrons by reflecting the 2450 keV beam from a deuterium oxide (D$_2$O) target. This allows the lowest nuclear recoil energies to be calibrated with decreased uncertainty. 

\section{Electronics and Controls}

\label{sec:FrontEnd}

The signal processing electronics, described in detail in Sec.~\ref{subsec:SignalFlow}, processes the signals from 494 TPC PMTs, 131 skin PMTs, and 120 outer-detector PMTs.
The electronics is designed to ensure a detection efficiency for single photoelectrons (PEs) of at least 90\%.   The PMT signals are digitized with a sampling frequency of 100~MHz and 14-bit accuracy.  The gain and shaping parameters of the amplifiers are adjusted to optimize the dynamic range for the PMTs.  The dynamic range for the TPC PMTs is defined by the requirement that the S2 signals associated with a full-energy deposition of the 164 keV $^{131m}$Xe activation line do not saturate the digitizers. Larger energy depositions will saturate a number of channels of the top PMT array, but the size of the S2 signal can be reconstructed using the S2 signals detected with the bottom PMT array.  The saturation of a few top PMTs for S2 signal will not impact the accuracy of the position reconstruction. For the skin PMTs, the dynamic range is defined by the requirement that the skin PMT signals associated with the interaction of a 511-keV $\gamma$-ray in the skin do not saturate the digitizers.  Simulations show that such an interaction can generate up to 200 PEs in a single PMT, depending on the location of the interaction. The dynamic range for the outer-detector PMTs is defined by the requirement that the size of the outer-detector PMT signals associated with neutron capture on Gd, which generates a $\gamma$-ray cascade with a total energy between 7.9 and 8.5~MeV, do not saturate the analog and digital electronics. Such events generate at most 100 PEs in a single PMT.  For muon interactions in the outer detector, a few PMTs may saturate, depending on the location of the muon track.  

The Data Acquisition system (DAQ) is designed to allow LED calibrations of the TPC PMTs in about 10 minutes. This requires an event rate of 4-kHz, resulting in a $\sim$340~Mb/s total waveform data rate. Monte Carlo simulations predict a total background rate is about 40~Hz. The background rate between zero and 40~keV is about 0.4~Hz. Due to the maximum drift time of 800~$\mu$s in the TPC, the rate for TPC source calibrations is limited to 150~Hz. A 150~Hz calibration rate results in a 10\% probability of detecting a second calibration event within the drift time of the previous calibration event. 

The slow control system is responsible for controlling and monitoring all LZ systems. It is described in detail in Sec.~\ref{subsec:Controls}.

\subsection{Signal Flow}
\label{subsec:SignalFlow}

\begin{figure}[h!]
\centering
\includegraphics[trim={0 0.0cm 0 0.0cm},clip,width=0.95\linewidth]{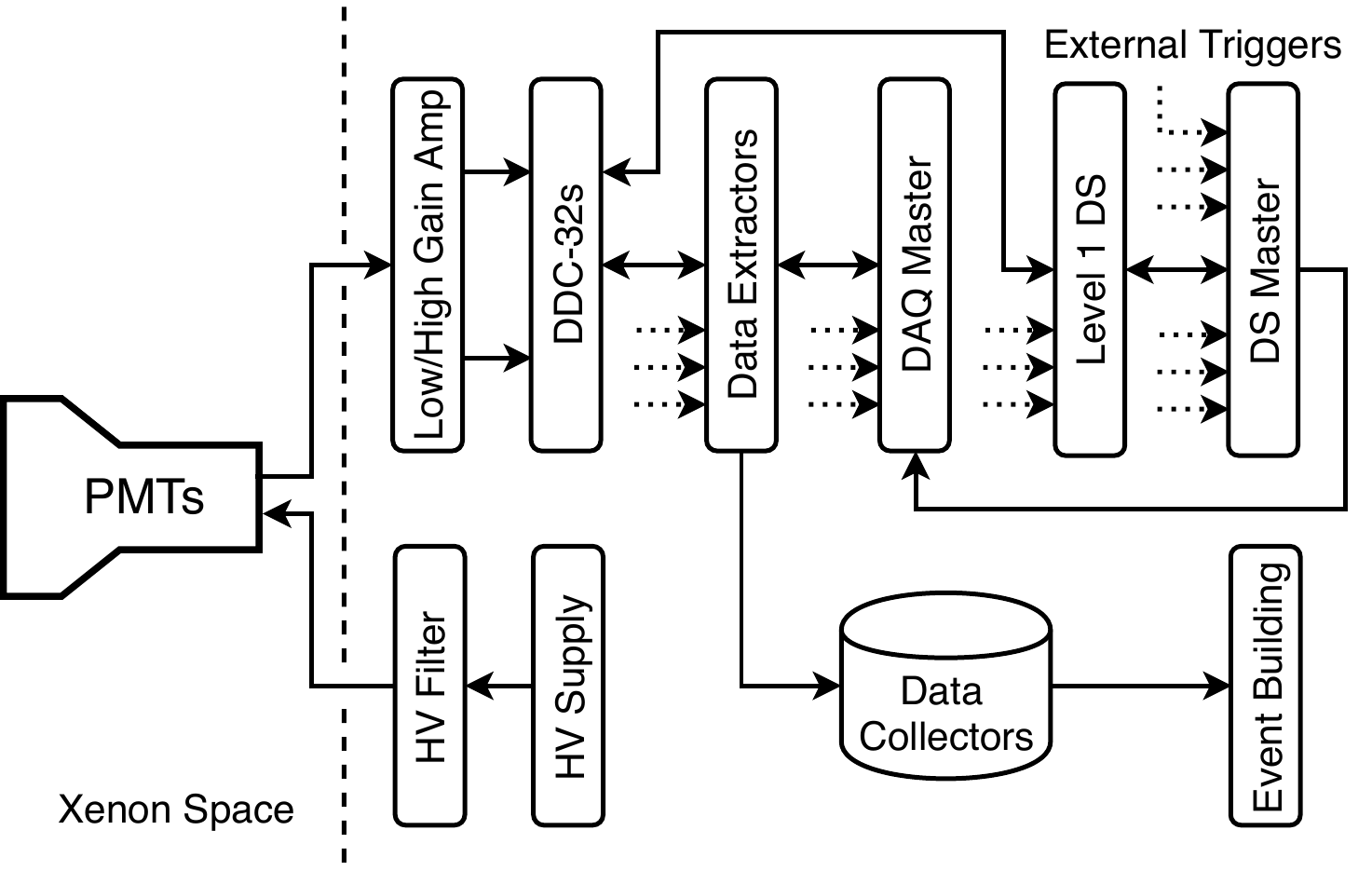}
\caption{Schematic of the signal processing of the TPC PMTs. The TPC and outer-detector PMTs use dual-gain signal processing. The skin PMTs only utilize the high-gain section of the amplifiers.  The signals are digitized using the DDC-32 digitizer, developed for LZ in collaboration with SkuTek~\cite{skutek}.  Event selection is made by the Data Sparsification (DS) system~\cite{druszkiewicz:2015pcl}.  }
\label{fig:signalFlow} 
\end{figure}

The processing of the signals generated by the TPC PMTs is schematically shown in Fig.~\ref{fig:signalFlow}. The TPC and skin PMTs operate at a negative HV, supplied by the HV system, using MPOD EDS 20130n\_504 and MPOD EDS 20130p\_504 HV distribution modules from from WIENER Power Electronics~\cite{wiener}. HV filters are installed at the HV flange on the breakout box. The PMT signals leave the breakout box via a different flange and are processed by the analog front-end electronics. The amplified and shaped signals are connected to the DAQ. The digitized data are sent to Data Collectors and stored on local disks. 
The PMTs of the outer-detector system operate at positive HV. The same type of amplifier used for the TPC and skin PMTs is also used for the outer-detector PMTs.

The PMT signals are processed with dual-gain amplifiers.  The low-gain channel has a pulse-area gain of four and a 30-ns full width at tenth maximum (FWTM) shaping-time constant. The high-gain channel has a pulse-area gain of 40 and a shaping time of 60-ns (FWTM). The high-gain channel is optimized for an excellent single PE response. The shaping times and gains are derived from one assumption: the DAQ has a usable dynamic range of 1.8~V at the input. A 0.2~V offset is applied to the digitizer channels in order to measure signal undershoots of up to 0.2~V.  Measurements with prototype electronics and the LZ PMTs have shown that the same amplifier parameters can be used for all of them.  For the TPC and the Outer Detector PMTs, both high-gain and low-gain channels are digitized; for the skin PMTs, only the high-gain channel is digitized.

The top-level architecture of the DAQ system is shown schematically in Fig.~\ref{fig:signalFlow}. The DDC-32s digitizers, eveloped for LZ in collaboration with SkuTek~\cite{skutek}, continuously digitize the incoming PMT signals and store them in circular buffers. When an interesting event is detected, the Data Extractors (DE) collect the information of interest from the DDC-32s. The DEs compress and stack the extracted data using their FPGAs and send the data to Data Collectors for temporary storage. The Event Builder takes the data organized by channels and assembles the buffers into full event structures for online and offline analysis. The DAQ operation is controlled by the DAQ Master for high-speed operations such as system synchronization and waveform selection, and by the DAQ Expert Control/Monitoring system, not shown in Fig.~\ref{fig:signalFlow}, for slow operations such as running setup/control and operator diagnostics. The entire system runs synchronously with one global clock. 

\begin{figure}[h!]
\centering
\includegraphics[trim={0 0.0cm 0 0.0cm},clip,width=0.95\linewidth]{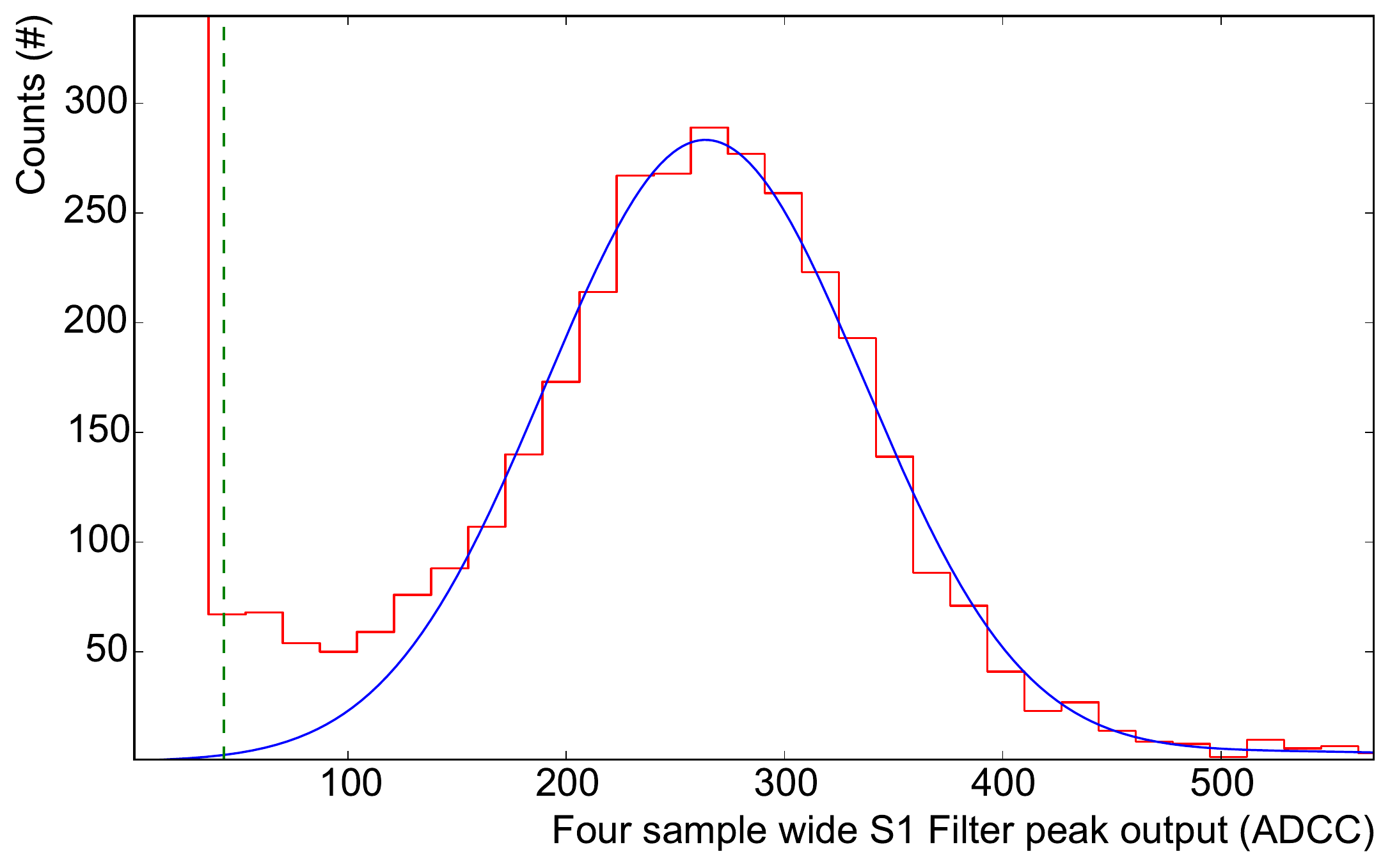}
\caption{Measurements of the TPC PMT response to a single PE.  The output of the four-sample wide S1 filter, in units of ADC Counts (ADCC), is shown.}
\label{fig:spheEff} 
\end{figure}

The performance of the entire signal processing chain has been evaluated in an electronics chain test.  Pre-production prototypes of the analog and digital as well as signal cables of the same type and length as those to be installed at SURF were used.  The measured response of a four-sample wide S1 filter is shown in Fig.~\ref{fig:spheEff}.  The measured single photoelectron (PE) efficiency is 99.8\%, much better than the requirement of 90\%.  The threshold at which the false-trigger rate is 1~Hz is 43 ADC Counts (ADCC) or 16\% of a single PE.

\subsection{Data Flow and Online Data Quality}
\label{subsec:DataFlow}

\begin{figure}[h!]
\centering
\includegraphics[trim={0 0.0cm 0 0.0cm},clip,width=0.95\linewidth]{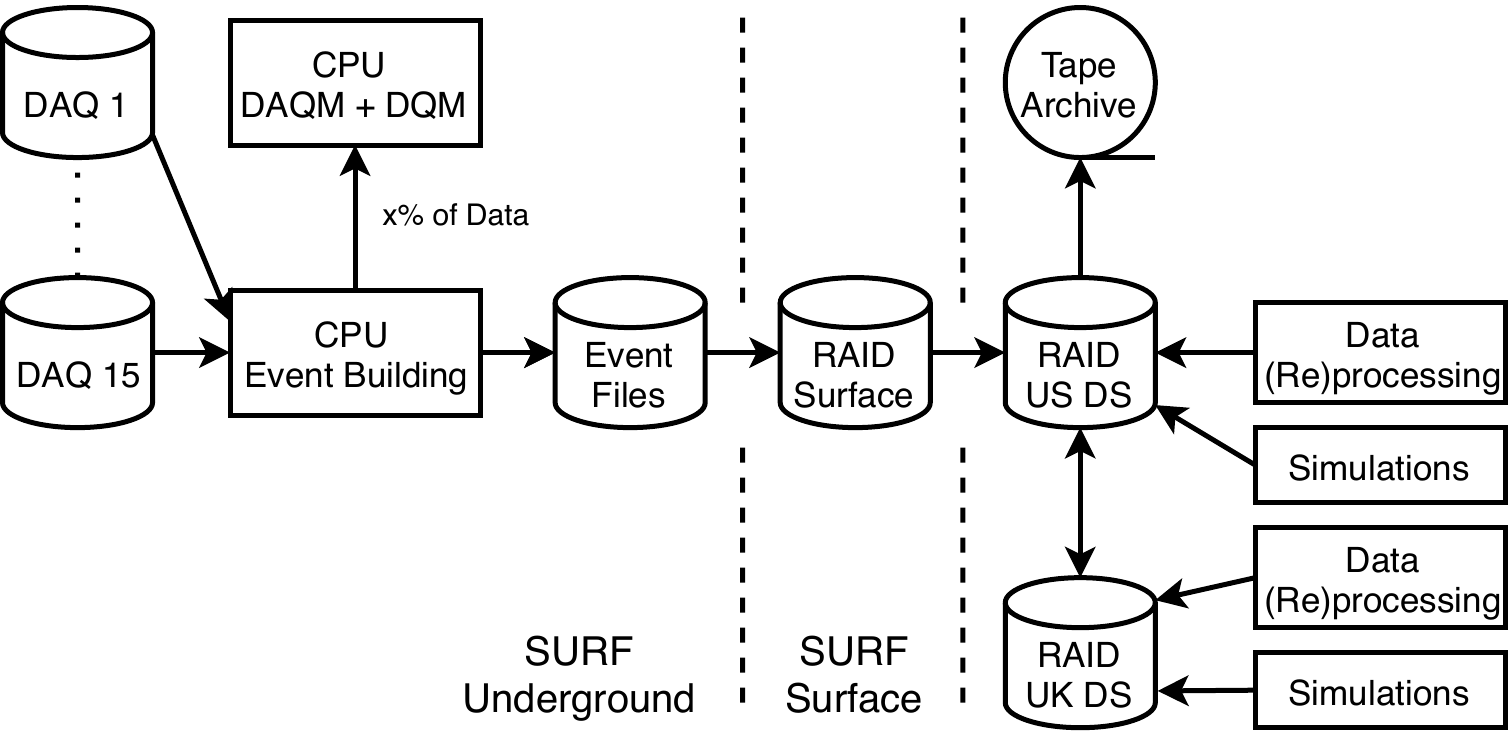}
\caption{A schematic of the LZ data flow.}
\label{fig:dataFlow} 
\end{figure}

The data flow is schematically shown in Fig.~\ref{fig:dataFlow}. Five event builders (EB) assemble the events by extracting the relevant information from the Data Collector disks, DAQ1-DAQ15. A 10 Gigabit per second (Gbs) line connects the Data Collectors to the Event Builders.  The event files are stored on the 16~TB local disk array of each EB, before being transferred to the 192~TB RAID array installed on the surface. From there, the event files are distributed to the data-processing centers for offline data processing and analysis.

Redundant connections exist between the surface and the Davis Cavern. A pair of managed 10~Gbs switches is installed on the surface and underground.  Each switch on the surface is connected via two fibers, travelling through the Yates and the Ross shafts, to the two underground switches.  Since each link supports 10~Gbs, this configuration can support data transfer rates of up to 40~Gbs.

A fraction of the data is analyzed online by the Detector Quality Monitor (DQM), running on a dedicated online server, installed in the Davis Laboratory. The DQM applies elements of the offline analysis to the data being collected, in order to monitor the performance of the detector.  For example, during $^{83m}$Kr calibrations, the DQM will monitor the electron life time and the energy resolution.  Various detector parameters, such as PMT multiplicity, hit distributions across both PMT arrays, and trigger rates, will be monitored. If significant deviations from prior observed patterns are seen, experts will be automatically notified via the slow control system. 

\subsection{Controls }
\label{subsec:Controls}

The Controls system performs supervisory control and monitoring of all the major subsystems of the experiment, including cryogenics, fluid handling, detector diagnostic sensors, high voltage, and electronics monitoring.  Not included in this system are the SURF-managed controls ensuring personnel safety (for example, oxygen deficiency alarms and sensors). The functionality provided by the Controls system can be classified in the following four categories: 
1) protection against xenon loss and contamination,
2) experiment parameter monitoring and logging,
3) control over LZ subsystems, and
4) providing the interface to operators.

In order to minimize risks associated with possible xenon loss or contamination, instruments and subsystems are divided in two major groups with respect to the perceived impact of their possible malfunction on the integrity of the xenon supply. Those where equipment failure or operator error can lead to xenon loss or contamination are designated ``critical" and the rest are ``non-critical". During a power failure, a combination of UPS and generator power will provide continuity to the critical components of the slow control system.  

The PLC system provides automatic protection in the case of an emergency. If the Xe pressure inside the ICV reaches a threshold value, or if there is an extended power outage in the Davis Cavern, the Xe compressors will activate and the vaporized xenon will be safely transferred to the Xe storage packs. In these scenarios Xe transfer occurs automatically without assistance from a human operator. Additionally, the PLC is programmed with a set of interlocks to protect the experiment from erroneous operator commands or equipment failure during routine operations.

\begin{figure*}[h!]
\centering
\includegraphics[width=0.85\linewidth]{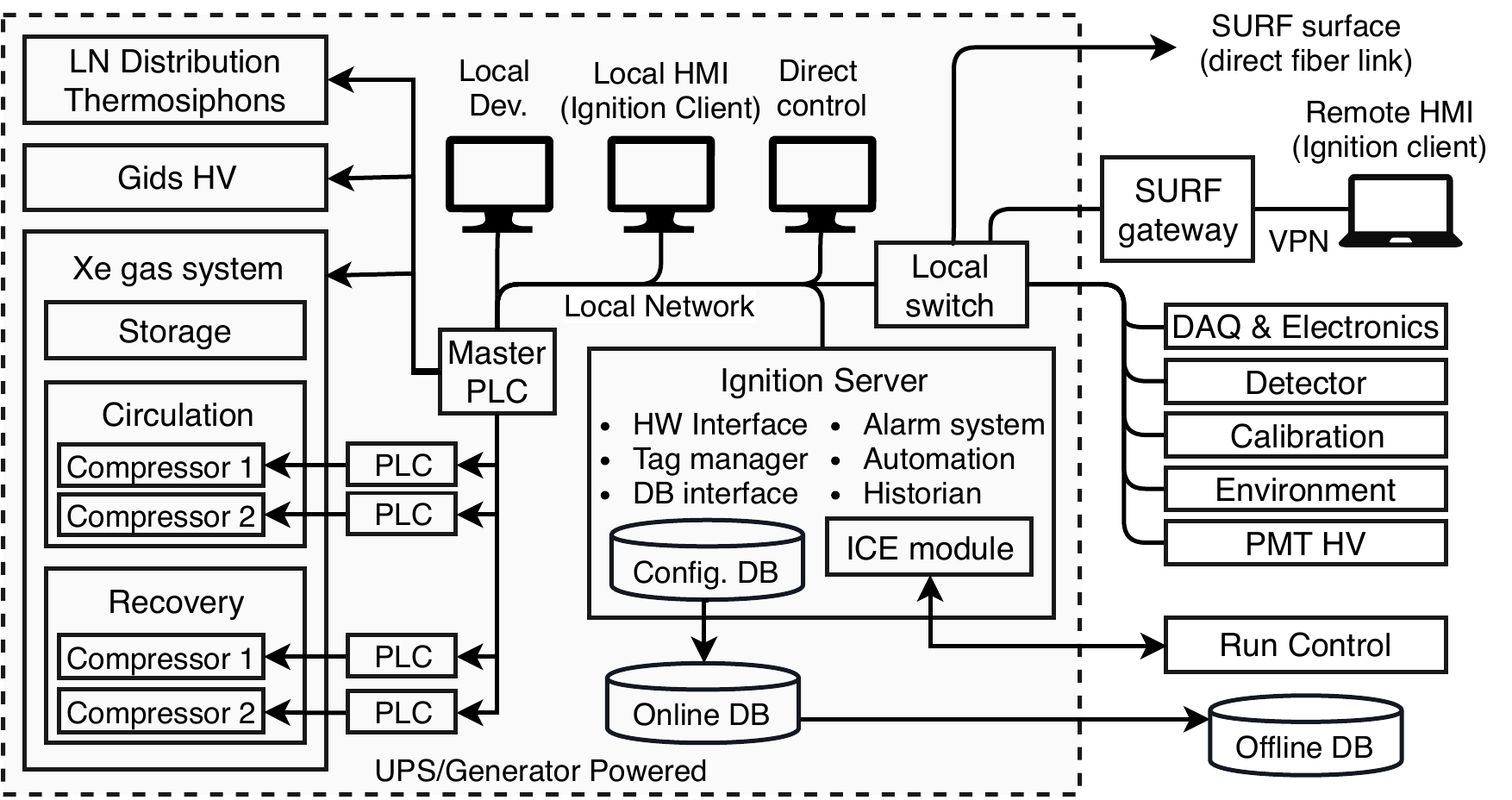}
\caption{Slow control functional diagram.}
\label{fig:SlowControl} 
\end{figure*}

The functional diagram of the slow control system and its interaction with the experiment subsystems and infrastructure is shown in Fig.~\ref{fig:SlowControl}. The core of the slow control system is composed of three components: (1) the integrated supervisory control and data acquisition (SCADA) software platform \emph{Ignition} from Inductive Automation~\cite{sqlbridge:2015}, (2) a Siemens SIMATIC S7-410H dual-CPU PLC with Redundant Hot Backup, allowing bumpless transfer from one active CPU to the backup, and (3) associated I/O modules~\cite{Siemens410-5h}.  Non-critical instruments connect directly to the Ignition server, typically using MODBUS-TCP over the slow control local network.  Critical instruments are managed by the PLCs, which in turn communicates with the Ignition server.  

The Ignition server provides a single operator interface, alarm system, and historical record for all slow control instrumentation. It provides authorized users with access to specific controls. In addition, the Ignition server provides the scripting engine for experiment automation. Ignition also provides a GUI for accessing historical data in the form of plots by local and remote clients. The configuration data (properties of sensors and controls, alarms, and user preferences) are stored in the local configuration database. 

For critical systems, which include xenon handling, cryogenics, and grid high voltage, PLCs add an additional layer between the slow control software and the physical instruments.  This ensures uninterrupted control of critical instruments and separates the low level logic governed by PLCs from the higher level operations run by the Ignition scripting engine.  

The PLC system is responsible for the automated recovery of xenon in emergency scenarios and must be robust against single-point failures.  This begins with the the Siemens dual-CPU system.  The system is powered by redundant 24~VDC supplies fed by two separate electrical panels, one of which is supported by an 8~kVA APC Symmetra UPS.  Both panels are also supported by backup diesel powered electrical generator, capable of powering the PLC system and xenon recovery compressors for the time required to complete xenon recovery.

Each Siemens CPU has its own connection to a common set of S7-300 I/O Modules, which connect directly to individual instruments/sensors, as well as to a set of PROFIBUS-DP Y-links, allowing for redundant control of PROFIBUS instruments.  In either case, redundancy does not in general extend to the I/O modules and instruments. However, those instruments critical for xenon recovery are duplicated to eliminate single-point failures.

Several integrated instruments, including the xenon compressors, vacuum pump systems, and the liquid nitrogen generator, have their own dedicated PLCs which function as PROFIBUS slaves to the Master PLC.  These smaller PLCs are either provided and programmed by the instrument vendor or, in the case of the xenon compressors, built and programmed by LZ.

Each of the four xenon compressors (two circulation and two recovery) is run by a dedicated Beckhoff CX8031 PLC, which governs compressor startup and shutdown sequences. The PLCs running the two xenon recovery compressors have the extra feature that they are capable of initiating xenon recovery in response to an over-pressure scenario, triggered by xenon pressure transducers connected directly to each PLC.  The intent is that emergency xenon recovery will be initiated and coordinated by the Master PLC, with this slave-initiated recovery as a backup.  

The choice of Ignition as the software platform for LZ Controls is based on the wide range of highly customizable, easy to use core functions and tools for development of efficient and robust SCADA systems. Ignition also comes with a versatile toolbox for GUI design and a comprehensive library of device drivers supporting most of the hardware used in LZ. One of these drivers supports Siemens PLCs allowing to export the PLC tags (internal variables). Devices connected to the PLC can be exposed to Ignition as sensors and controls. For the rest of the devices the preferred protocol is MODBUS which is also supported by the Ignition server. For those devices that do not support MODBUS (e.g. RGA units), two interfacing methods are envisaged: 1) custom Java drivers written in the Ignition software development kit, and 2) Python MODBUS server with plugin system or custom Python drivers.

The operator interface is comprised of a set of panels, organized into a tree view. At a lower GUI level, the panels represent the experiment subsystem via functional diagrams with relevant information on the state of sensors and controls displayed in real time. The authorized system experts can also use these GUI panels to alter the controls not protected by the PLC interlocks. At a higher level, several panels show the real-time status of the system and subsystems from which a trained operator is able to tell the overall health of the system. These panels also allow to assess high level information, for example alarm status, current operation mode, status of automation scripts, etc. At a very high level, a single summary plot of the entire system can be prepared by slow control and sent to run control for display in a single status panel on the run control GUI.

\section{Detector Assembly}
\label{sec:UGAssembly}

LZ is being installed inside the existing 25' diameter, 20' 
tall Davis Cavern water tank used for the LUX experiment.
Access to the water tank is down the 4850' vertical 
Yates shaft and through horizontal drifts. 
Some large items of equipment are segmented and 
transported underground in sections, including the 
OCV (three sections), and the tall AVs (four sections).
Other items, like the ICV and its TPC payload, and the 
LXe tower, are transported and installed after being 
fully assembled on the surface.

Substantial changes to the infrastructure of the Davis Campus were 
required to accommodate the detector size increase from LUX to LZ.  This included 
creating more floor space with a platform for cable breakouts, 
conversion of the compressor room roof to a space for purification and 
Xe sampling equipment, and converting two previously unused excavations to 
spaces for Xe storage and Rn removal equipment.

Underground assembly started in 2018 with the transport of the 12-foot 
tall AVs. Each AV was contained in a heavy steel frame. 
At the top of the Yates shaft the frame was mounted 
in a rotatable assembly, 
slung under the cage, and lowered to the 4850' level. 
The rotatable assembly was used to move each 5000 pound unit around 
duct work and other obstacles between the cage and the 
Davis Cavern upper deck.  
Each AV was wrapped in two plastic bags to keep dust from contaminating 
the acrylic and the water tank. The AVs 
are sensitive to large temperature 
changes so timing and speed were also important 
considerations during transport.

\begin{figure}[t!]
\centering
\includegraphics[trim={0 0.0cm 0 0.0cm},clip,width=0.90\linewidth]{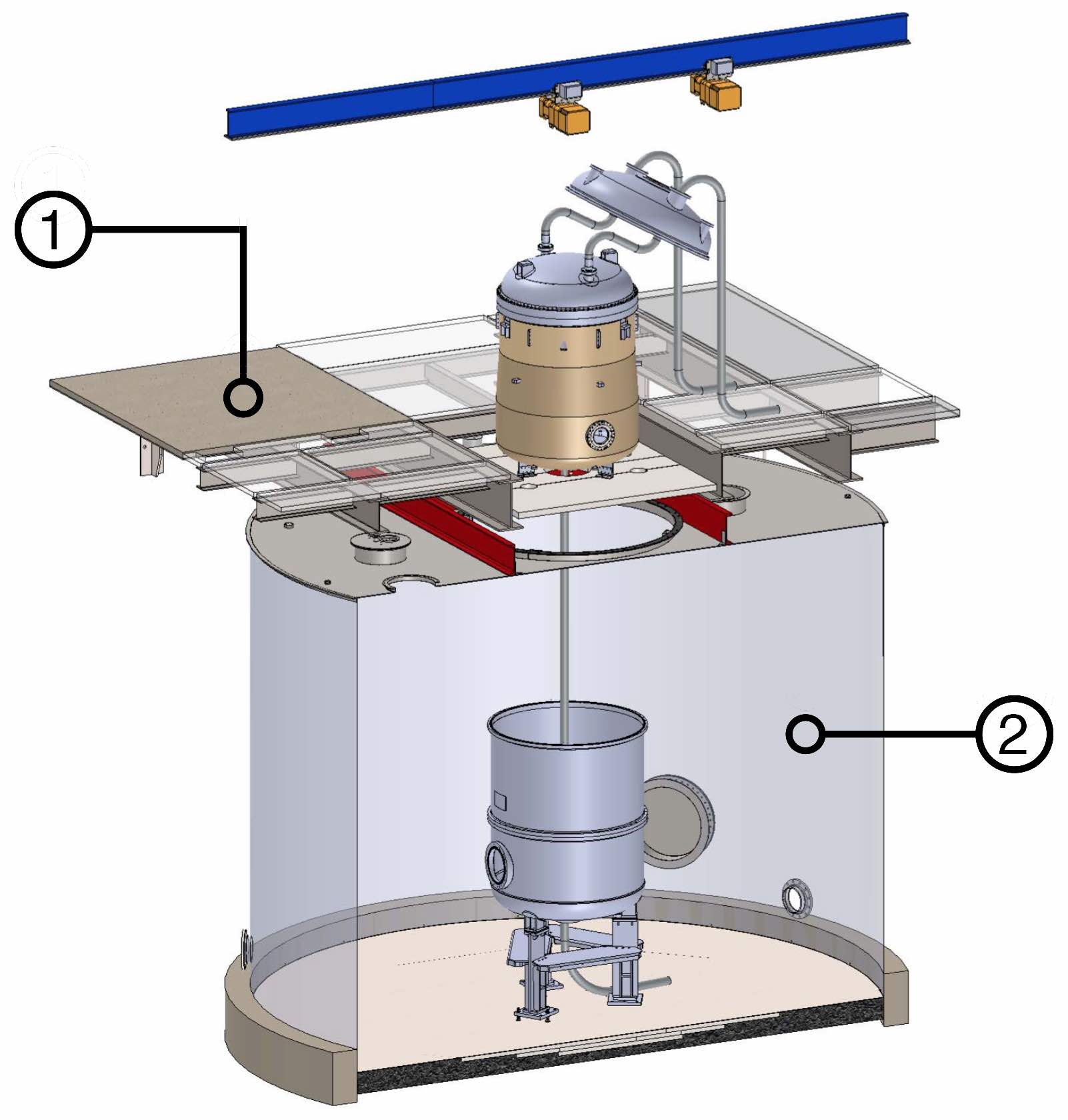}
\caption{Intermediate assembly. A section through the Davis Campus showing the OCV ready to receive the ICV, and the ICV suspended above with cable conduits. 1-Deck; 2-Water tank.}
\label{fig:intermediateassembly} 
\end{figure}

After the four tall AVs were placed inside 
the water tank, the three sections
of the OCV were transported underground and installed on  
support legs inside. After a leak cheak, and 
a final washing of the water tank and its equipment, 
the OCV top was removed and stored to prepare for ICV installation, 
and cleanroom protocols were instituted for entering and exiting the area. 
An underground radon reduction system treats the air supplied 
to the water tank during the remainder of the assembly.

Much of the TPC assembly work was done in the Surface Assembly Lab 
(SAL) at SURF because of its superior logistics 
compared to the underground.  
A well sealed class 1000 clean 
room with reduced radon air supply~\cite{ateko} 
and a clean monorail crane was used 
for cryostat acceptance, TPC assembly, and insertion of the TPC into the 
ICV. The Reduced Radon System (RRS) has been 
measured to produce air laden with
radon at a specific activity of 4~mBq/m$^3$ given 
input air at about 8.5~Bq/m$^3$.
The PMT and instrumentation cables are routed through 
three reinforced bellows attached
to the ICV, and the entire assembly is transported underground
together. During transport the bellows are sealed 
and the ICV is purged with boil off LN. This 
begins the process of removing H$_2$O, O$_2$, and Kr from 
the large mass of PTFE in the detector. All subsequent assembly
steps are designed to keep the TPC under nitrogen purge until it 
can be evacuated and filled with Xe gas.

The TPC is assembled and installed into the ICV vertically. 
To move underground, the assembly is rotated 
to horizontal, set in a transport frame, 
and moved to the Yates headframe.
Slings attached to the underside of the cage 
are used to lift the assembly into a vertical position 
underneath the cage. 
This process is reversed at the 4850' level,  
returning it to horizontal, for transport down the drift.
It is set vertical again
on a shielding platform over the water tank.  A temporary clean room is 
constructed around the ICV, and the last protective bag is removed.
The ICV is connected to the 
OCV top by three tie bars (instrumented threaded rods) that allow for 
precision leveling of the TPC relative to the liquid Xe surface while 
minimizing thermal losses. During final lowering into the 
OCV the ICV is supported by the tie bars.
Once sealed, the ICV is evacuated, and the AVs
and OD PMTs are assembled around the OCV in the water tank. 
A rendering of the ICV just prior to nesting within the OCV is shown in Fig.~\ref{fig:intermediateassembly}.

A second radon reduction system is available in the Davis Cavern to supply
air depleted of radon. It has demonstrated an output of 100~mBq/m$^3$ 
given input air of 70~Bq/m$^3$. 

The water and GdLS are
co-filled to minimize stress in the AV 
walls.  The GdLS will be transported underground in 150 sealed 
barrels. Each is placed on a scale in a clean tent before connection to 
the liquid transfer system.  The liquid is pumped into the GdLS 
reservoir and distributed to the AVs.  Once the liquids are filled 
detector commissioning can begin.

\section{Materials Screening \& Backgrounds}
\label{sec:Materials}

Material screening is the primary route to controlling the ER and NR backgrounds resulting from radioactivity in the experiment. Measurements of radioactive nuclides in and on all detector components are required. The ubiquitous and naturally occurring radioactive materials (NORM) of particular concern are the \Pgg-ray emitting nuclides $^{40}$K; $^{137}$Cs; and $^{60}$Co, as well as $^{238}$U, $^{235}$U, $^{232}$Th, and their progeny. The U and Th chains are also responsible for neutron production following spontaneous fission and ($\alpha$,n) reactions. Kr and Rn outgassing from materials into the Xe also results in ER backgrounds, and $\alpha$-emitting Rn daughters can contribute to neutron backgrounds when deposited on certain materials.

Fixed contamination, referring to non-mobile NORM nuclides embedded within materials, is the dominant source of neutron and $\gamma$-ray emission in LZ. To ensure a sub-dominant contribution from fixed contaminants, relative to irreducible backgrounds, all materials considered for use in the construction of the experiment are screened for NORM nuclides to $\approx$0.2~mBq/kg sensitivities, equivalent to tens of parts per trillion (ppt) g/g for $^{238}$U and $^{232}$Th. This ensures a maximum  contribution from fixed contamination of less than 0.4~NR and $1\times 10^{-6}$ events/(kev $\cdot$ kg $\cdot$ year) ER in the LZ exposure. For materials such as PTFE, which are produced in granular form before being sintered in molds, plate-out constitutes additional risk because surface contamination of the granular form becomes contamination in bulk when the granules are poured into molds. A limit of 10~mBq/kg of $^{210}$Po and $^{210}$Pb in PTFE maintains an NR contribution of $<$0.1 count in the LZ exposure. 

Radon emanated from within detector components is the dominant contributor to background in LZ, primarily through the ``naked'' beta emission from $^{214}$Pb in the $^{222}$Rn sub-chain as it decays to $^{214}$Bi. The $^{214}$Bi beta decay itself is readily identified by the subsequent $^{214}$Po alpha decay that would be observed within an LZ event timeline (T$_{1/2}$=160~$\mu$s). Similar coincidence rejection also occurs where beta decay is accompanied by a high-energy $\gamma$-ray, which may still be tagged by the Xe Skin or OD vetoes even if it leaves the active Xe volume. $^{220}$Rn generates $^{212}$Pb, resulting in $^{212}$Bi-$^{212}$Po sequential events which can be tagged. Radon daughters are readily identified through their alpha decay signatures and can be used to characterize the $^{222}$Rn and $^{220}$Rn decay chain rates and distributions in the active region, providing a useful complement to estimating radon concentration from the beta decay contribution to the ER background.  

The specific activity of $^{222}$Rn in LZ is required to be less than 2~$\si{\micro}$Bq/kg of LXe, equivalent to 20~mBq. All xenon-wetted components in LZ have been constructed from low-radon materials selected through dedicated measurement facilities. Due to the large number of expected emitters, these facilities were required to achieve sensitivity of 0.2~mBq to $^{222}$Rn. Measurements are made at room temperature, however, the expected emanation can depend strongly on temperature depending on the source material. A conservative approach is adopted in estimating radon emanation in our model, taking credit for a reduction at LXe temperatures wherever this is supported in the literature. Significant background from $^{220}$Rn is not expected given its very short half-life; we conservatively include in our background model a contribution from $^{220}$Rn of 20\% of the ER counts from $^{222}$Rn.  

The accumulation of $^{222}$Rn-daughters plated-out during the manufacture and assembly of components as well as dust and particulates contribute to the LZ background. The $\alpha$-particle emitting Rn-daughters can induce neutrons via ($\alpha$,n) processes, particularly problematic for materials with large cross-sections for this process such as flourine, present in the TPC walls (PTFE). Plate-out on the inside of the TPC walls causes $\alpha$-particles and recoiling ions to enter the active volume. The risk of mis-reconstructing these recoils into the fiducial volume in particular sets stringent constraints on plate-out on the PTFE. LZ instituted a target for plate-out of $^{210}$Pb and $^{210}$Po of less than 0.5~mBq/m$^{2}$ on the TPC walls and below 10~mBq/m$^{2}$ everywhere else. 

Generic dust containing NORM also releases $\gamma$-rays and induce neutron emission. Dust is further expected to be the single largest contributor to radon emanation. Dust contamination is limited to less than 500~ng/cm$^{2}$ on all wetted surfaces in the detector and xenon circulation system. Under the conservative assumption that 25\% of $^{222}$Rn is released from the dust, via either emanation or recoils out of small grain-size particulates, this limits the $^{222}$Rn activity from dust to less than 10 mBq. 

\renewcommand{\arraystretch}{1.8}
\begin{table*}[t]
\footnotesize
\caption{Primary material radio-assay techniques, indicating isotopic sensitivity and detection limits, as well as typical throughput or single-sample measurement duration.}
\centering
\begin{tabular}{ |>{\centering\arraybackslash}m{1.8cm} | >{\centering\arraybackslash}m{1.7cm}|>{\centering\arraybackslash}m{2.1cm}|>{\centering\arraybackslash}m{1.0cm}|>{\centering\arraybackslash}m{1.4cm}|>{\centering\arraybackslash}m{2.8cm}|>{\centering\arraybackslash}m{2.6cm}|>{\centering\arraybackslash}m{1.4cm}|}
\hline
\textbf{Technique} & \textbf{Isotopic Sensitivity} & \textbf{Typical Sensitivity} & \textbf{Sample \newline Mass} & \textbf{Sampling Duration} & \textbf{Destructive/Non-destructive and Notes} & \textbf{Locations (and Number of Systems if $>1$)} & \textbf{Samples Assayed}\\
\hline

\textbf{HPGe} & 
 $^{238}$U, $^{235}$U, $^{232}$Th chains, $^{40}$K, $^{60}$Co, $^{137}$Cs any $\gamma$-ray emitter  & 
\SI{5e-11}{g/g}~U, \SI{e-10}{g/g}~Th &
\si{\kg} & 
Up to 2 weeks &
Non-destructive, very versatile, not as sensitive as other techniques, large samples & SURF~$\times6$, LBNL~$\times1$, U.~Alabama~$\times2$, Boulby~$\times7$
& 926 \\
\hline
\textbf{ICP-MS} & 
 $^{238}$U, $^{235}$U, $^{232}$Th (top of chain) & 
\SI{e-12}{g/g}& \si{mg} to \si{g} & 
Days &
Destructive, requires sample digestion, preparation critical & UCL, IBS, BHUC, U.~Alabama
& 157\\
\hline
\textbf{NAA} & 
 $^{238}$U, $^{235}$U, $^{232}$Th (top of chain), K & 
\SIrange{e-12}{e-14}{g/g} & 
\si{g} &
Days to weeks & 
Destructive, useful for non-metals, minimal sample preparation & Irradiated at MITR-II, HPGe assay at U.~ Alabama
&  3\\
\hline
\textbf{GD-MS} & 
 $^{238}$U, $^{235}$U, $^{232}$Th (top of chain) &  \SI{e-10}{g/g} & 
\si{mg} to \si{g} &
Days & 
Destructive, minimal matrix effects, cannot analyze ceramics and other insulators & National Research Council Canada
& 2 \\
\hline
\textbf{Radon Emanation} & 
$^{222}$Rn & % Removed 220Rn since we do not screen for this, C.Hall 3/13/19 
\SI{0.1}{mBq} & 
\si{\kg} & 
1 to 3 weeks &
Non-destructive, large samples, limited by size of emanation chamber & UCL~$\times2$, U.~Maryland, SDSM\&T~$\times2$, U.~Alabama~$\times2$
& 175 \\
\hline
\textbf{Surface $\alpha$}& $^{210}$Pb, $^{210}$Bi, $^{210}$Po & 120~$\alpha$/(m$^{2} \cdot$ day) & g to kg & $<$1 week & Non-destructive, thin samples, large surface area required & SDSM\&T~(Si), Brown~(XIA), Boulby~(XIA), U.~Alabama~(Si) 
& 306 \\[1ex] \hline
\end{tabular}
\label{table:nonlin}
\end{table*}

\subsection{HPGe + MS Techniques and Instruments}
\label{sec:HPGeMS}

The LZ screening campaign deploys several mature techniques for the identification and characterization of radioactive species within these bulk detector materials, primarily \Pgg-ray spectroscopy with High Purity Germanium (HPGe) detectors and Inductively-Coupled Plasma Mass Spectrometry (ICP-MS), supported by Neutron Activation Analysis (NAA). These complementary techniques collectively produce a complete picture of the fixed radiological contaminants.

Sensitivity to U and Th decay chain species down to $\approx$10~ppt has been demonstrated using ultralow-background HPGe detectors. HPGe can also assay $^{60}$Co, $^{40}$K, and other radioactive species emitting $\gamma$-rays. This technique is nondestructive and, in addition to screening of candidate materials, finished components can be assayed prior to installation. Under the assumption of secular equilibrium, the U and Th content, assuming natural terrestrial abundance ratios, may be inferred from the measurement of isotopic decay emissions lower in their respective chains. However, secular equilibrium can be broken through removal of reactive nuclides during chemical processing or through emanation. HPGe readily identifies the concentrations of nuclides from mid- to late-chain $^{238}$U and $^{232}$Th, particularly those with energies in excess of several hundred keV. Background-subtracted $\gamma$-ray counting is performed around specific energy ranges to identify radioactive nuclides. Taking into account the detector efficiency at that energy for the specific sample geometry allows calculation of isotopic concentrations. A typical assay lasts 1--2 weeks per sample to accrue statistics at the sensitivities required for the LZ assays. These direct $\gamma$-ray assays probe radioactivity from the bulk of materials and may identify equilibrium states. 

Sixteen HPGe detectors located in facilities both above- and underground have been used for LZ, with differences in detector types and shielding configuration providing useful dynamic range both in terms of sensitivity to particular nuclides and to varying sample geometries. The detectors are typically several hundreds of grams to several kilograms in mass, with a mixture of n-type, p-type, and broad energy Ge (BEGe) crystals, providing relative efficiency at the tens of percent through to in excess of 100\% (as compared to the detection efficiency of a ($3\times3$)-inch NaI crystal for 1.33~MeV $\gamma$-rays from a $^{60}$Co source placed 25~cm from the detector face). While p-type crystals can be grown to larger sizes and hence require less counting time due to their high efficiency, the low energy performance of the n-type and broad energy crystals is superior due to less intervening material between source and active Ge. Clean samples are placed close to the Ge crystal and assayed for several days to weeks in order to accrue sufficient statistics, depending on the minimum detectable activity (MDA). The detectors are generally shielded with low-activity Pb and Cu, flushed with dry nitrogen to displace the Rn-carrying air, and sometimes are surrounded by veto detectors to suppress background from Compton scattering that dominates the MDA for low-energy $\gamma$-rays. To reduce backgrounds further, most of the detectors are operated in underground sites~\cite{Scovell:2017srl,Mount:2017iam}, lowering the muon flux by several orders of magnitude. We also utilize a number of surface counters, some of them employing active cosmic rate veto systems, that are particularly useful for pre-screenings before more sensitive underground assays. 

To ensure uniform analysis outputs for all HPGe detectors, a cross-calibration program was performed using all detectors active in 2014. This involved the blind assay of a Marinelli beaker containing $\approx2$~kg of Rhyolite sourced from the Davis Cavern at SURF. This sample had previously been characterized using the MAEVE p-type HPGe detector at LBNL. Across the eight detectors online at the time, assays for both $^{238}$U and $^{232}$Th were within $1\sigma$ of each-other. As additional detectors have been brought online, consistency has been assured by cross-calibration intra-facility.

ICP-MS offers precise determination of elemental contamination with potentially up to 100$\times$ better sensitivity for the progenitor U and Th concentrations compared to \Pgg-ray spectroscopy. Since ICP-MS directly assays the $^{238}$U, $^{235}$U, and $^{232}$Th progenitor activity it informs the contribution to neutron flux from ($\alpha$,n) in low-Z materials, but also the contribution from spontaneous fission, which in specific materials can dominate. However, it cannot identify daughter nuclides in the U and Th decay chains that are better probed by HPGe. The ICP-MS technique assays very small samples that are atomized and measured with a mass spectrometer. As a destructive technique, it is not used on finished components. The limitation of ICP-MS is that the sample must be acid soluble and that several samples from materials must be screened to probe contamination distribution and homogeneity. Assays take 1--2 days per material, dominated by the sample preparation time, where extreme care must be taken to avoid contamination of solvents and reactants. 

ICP-MS assays for LZ materials have been performed using several facilities, the majority of which operate Agilent 7900 ICP-MS systems within minimum ISO Class 6 cleanrooms~\cite{Dobson:2017esw}. These are capable of achieving sensitivity to U and Th in materials at the level of several ppt. Protocols and methodologies for sample preparation are largely based on well established procedures~\cite{Leonard:2007uv,LaFerriere:2014rva,Grinberg:2005}. 

These measurements take significant time owing to the need for high statistics, standard addition calibration of high-concentration samples, and frequent machine cleaning. For more routine measurements, the backgrounds are simply monitored and reported as an equivalent concentration, and lower concentration samples allow for external calibrations and a significantly relaxed cleaning schedule, allowing to measure less demanding samples at the rate of several per day with sensitivities to U and Th on the order of a few hundred ppt.  Finally, some initial work has been done to develop measurements of potassium using the cold plasma configuration, leading to measurements of a few hundred ppb of K.

NAA has been used by LZ to assay PTFE to sub-ppt g/g levels of $^{238}$U and $^{232}$Th that is not well suited to HPGe, due to the sensitivity, nor ICP-MS, due to difficulty in digesting PTFE. However, as with ICP-MS, this technique requires small sample masses, does not assay finished components, and assumptions of secular equilibrium need to be made since this technique measures the top of the U and Th chains. Samples are irradiated with neutrons from a reactor to activate some of the stable nuclides, which subsequently emit $\gamma$-rays of well-known energy and half-life that are detected through $\gamma$-ray spectroscopy. Elemental concentrations are then inferred, using tabulated neutron-capture cross sections convolved with the reactor neutron spectra. Depending on the surface treatment, NAA can probe both bulk and surface contamination. Its application and sensitivity are limited by the composition of the material.

\subsection{Radon Emanation Techniques and Instruments}

Radon emanation measurements of all xenon-wetted components within the inner cryostat and those in the gas system that come into contact with Xe during experimental operation have been performed using four facilities available to LZ, listed in Table~\ref{table:nonlin}. In all of the facilities radon is accumulated in emanation chambers, where samples typically remain for two weeks or more. In many cases multiple measurements are performed per sample. The radon is then transferred using a carrier gas to a detector to be counted. 

In one of the stations, the radon atoms are collected and counted by passing the radon-bearing gas through liquid, dissolving most of the radon. The scintillator is then used to count Bi-Po coincidences, detected through gated coincidence logic, using one PMT viewing the scintillator. In the other three stations, radon atoms and daughters are collected electrostatically onto silicon PIN diode detectors to detect $^{218}$Po and $^{214}$Po alpha decays. The radon screening systems have been developed such that anything from individual components through to sections of LZ pipework may be assayed.

All stations were initially evaluated using calibrated sources of radon, with a cross-calibration program performed to ensure the accuracy of each system's overall efficiency and ability to estimate and subtract backgrounds. 

The radon-emanation screening campaign extends beyond the material selection and construction phase and into detector integration and commissioning phases. A system is available underground at SURF in order to screen large-scale assembled detector elements and plumbing lines. As pieces or sections are completed during installation of gas pipework for the LZ experiment, they are isolated and assessed for Rn emanation and outgassing for early identification of problematic seals or components that require replacement, cleaning, or correction. 

\subsection{Surface Assays (XIA, Si, dust microscopy) }

Two sensitive detectors have been used to carry out assays of Rn plate-out to ensure the requirements are met and inform the experiment background model. The first is the commercial XIA Ultralo-1800 surface alpha detector system, suitable for routine screening of small samples including witness plates and coupons deployed during component assembly and transport to track exposure. The second detector employs a panel of large-area Si detectors installed in a large vacuum chamber. These systems exceed the requisite sensitivity to $^{210}$Po at the level of 0.5~mBq/m$^{2}$. Dust assays are performed using high-powered microscopy and x-ray fluorescence techniques.

\renewcommand{\arraystretch}{1.4}
\begin{table*}[]
\centering
\footnotesize
\label{table:backgrounds_condensed}
\caption[BackgroundsTable]{The estimated backgrounds from all significant sources in the LZ 1000~day WIMP search exposure. Mass-weighted average activities are shown for composite materials. Solar $^{8}B$ and hep neutrinos are only expected to contribute at very low energies (\textit{i.e. WIMP masses}) and are excluded from the table. 
}
\begin{tabular}{| m{12.6cm} | >{\centering\arraybackslash}m{0.7cm} | >{\centering\arraybackslash}m{0.7cm} |}
\hline
\multirow{ 2}{*}{\textbf{Background Source}} & \textbf{ER} & \textbf{NR} \\
& \textbf{(cts)} & \textbf{(cts)} \\
\hline \hline
\textbf{Detector Components} & \textbf{9}  & \textbf{0.07}    \\
\hline
\hline
{\textbf{Surface Contamination}}			& \textbf{40}   & \textbf{0.39} \\
\hline
{Dust (intrinsic activity, 500 ng/cm$^{2}$)}& 0.2  & 0.05 \\
{Plate-out (PTFE panels, 50 nBq/cm$^{2}$)}	& -    & 0.05 \\
{$^{210}$Bi mobility (0.1 $\mu$Bq/kg)}		& 40.0 & -	  \\
{Ion misreconstruction (50 nBq/cm$^{2}$)}	& -    & 0.16 \\
{$^{210}$Pb (in bulk PTFE, 10 mBq/kg)}		& -	   & 0.12 \\
\hline \hline
{\textbf{Laboratory and Cosmogenics}}		& \textbf{5}    & \textbf{0.06} \\
\hline
{Laboratory Rock Walls}						& 4.6  & 0.00 \\
{Muon Induced Neutrons}						& -    & 0.06 \\
{Cosmogenic Activation}						& 0.2  & -    \\
\hline \hline
{\textbf{Xenon Contaminants}}				& \textbf{819}  & \textbf{0} \\
\hline
{$^{222}$Rn (1.81 $\mu$Bq/kg)}			 	& 681  & - \\
{$^{220}$Rn (0.09 $\mu$Bq/kg)} 				& 111  & - \\
{$^{nat}$Kr (0.015 ppt)}		       		& 24.5 & - \\
{$^{nat}$Ar (0.45 ppb)} 	       			& 2.5  & - \\
\hline \hline
{\textbf{Physics}}							& \textbf{322}  & \textbf{0.51} \\
\hline
{$^{136}$Xe 2$\nu\beta\beta$}               & 67   & - \\
{Solar Neutrinos: $pp$+$^{7}$Be+$^{13}$N} 	& 255  & - \\
{Diffuse Supernova Neutrinos}	          	& -    & 0.05 \\
{Atmospheric Neutrinos}		            	& -    & 0.46 \\
\hline \hline
{Total}                                     & 1,195 & 1.03 \\
{Total (with 99.5~$\%$ ER discrimination, 50~$\%$ NR efficiency)} & 5.97 & 0.51 \\
\hline
{\textbf{Sum of ER and NR in LZ for 1000 days, 5.6 tonne FV, with all analysis cuts}} & \multicolumn{2}{c|}{\textbf{6.49}} \\
\hline
\end{tabular}
\end{table*}

\subsection{Cleaning procedures and protocols (ASTM standards)}

A rigorous program of cleanliness management is implemented to ensure that the accumulated surface and dust contamination are monitored, tracked and do not exceed requirements. All detector components that contact xenon must be cleaned and assembled according to validated cleanliness protocols to achieve the dust deposition levels below 500~ng/cm$^{2}$ and plate-out levels below 0.5~mBq/m$^{2}$ for the TPC inner-walls, and 10~mBq/m$^{2}$ everywhere else. Witness plates accompany the production and assembly of all detector components to ensure QC and demonstrate QA through the plate-out and dust assays. The titanium 
cryostat was cleaned by AstroPak~\cite{Astropak} to 
ASTM standard IEST-STD-CC1246 (rev E) 
level 50R1 (ICV) and level VC-0.5-1000-500UV (OCV). 
The ICV cleaning standard is equivalent to the requirement 
that mass density of dust be less than 100~ng/cm$^2$.
The vessels were etched according to ASTM B-600.

As described in Sec.~\ref{sec:UGAssembly}, detector integration is done in a reduced-radon cleanroom at the SAL at SURF. Dust and plate-out monitoring on-site was continuously performed to measure and maintain compliance with tolerable dust and plate-out levels. 

\subsection{Backgrounds summary }

Measured material radioactivity and anticipated levels of dispersed and surface radioactivity are combined with the Monte Carlo simulations and analysis cuts to determine background rates in the detector. Table~\ref{table:backgrounds_condensed} presents integrated background ER and NR counts in the 5.6~tonne fiducial mass for a 1000 live day run using a reference cut-and-count analysis, both before and after ER discrimination cuts are applied.
For the purposes of tracking material radioactivity throughout the design and construction of LZ, Table~\ref{table:backgrounds_condensed} is based on a restricted region of interest relevant to a 40~GeV/c$^{2}$ WIMP spectrum, equivalent to approximately 1.5--6.5~keV for ERs and 6--30~keV for NRs. 

The expected total from all ER(NR) background sources is 1195(1.24) counts in the full 1000~live day exposure. Applying discrimination against ER at 99.5\% for an NR acceptance of 50\% (met for all WIMP masses given the nominal drift field and light collection efficiency in LZ~\cite{Mount:2017qzi}) suppresses the ER(NR) background to 5.97(0.62) counts. Radon presents the largest contribution to the total number of events. Atmospheric neutrinos are the largest contributor to NR counts, showing that LZ is approaching the irreducible background from coherent neutrino scattering~\cite{billard:2013qya}. 

\section{Offline Computing} \label{sec:Offline}

The LZ data is stored, processed and distributed using two data centers, one in the U.S. 
and one in the U.K. Both data centers are capable of storing, processing, simulating and 
analyzing the LZ data in near real-time. Resource optimization, redundancy, and ease of 
data access for all LZ collaborators are the guiding principles of this dual data-center design. 

LZ raw data are initially written to the surface staging computer at SURF, which was designed
with sufficient capacity to store approximately two months of LZ data, running in WIMP-search
mode. This guarantees continuity of operations in case of a major network outage. The surface 
staging computer at SURF transfers the raw data files to the U.S. data center, where initial 
processing is performed. The reconstructed data files are made available to all groups in the 
collaboration and represent the primary input for the physics analyses. The  U.S. data center 
is hosted at the National Energy Research Scientific Computing (NERSC) center~\cite{NERSC:web}. 
Data simulation and reconstruction at NERSC is performed using a number of Cray 
supercomputers~\cite{Perlmutter:web}.

The raw and reconstructed files are mirrored to the U.K. data center (hosted at Imperial College London)
both as a backup, and to share the load of file access and processing. Subsequent 
reprocessing of the data (following new calibrations, updated reconstruction and identification 
algorithms, etc.) is expected to take place at one or both locations, with the newly generated 
files mirrored at both sites and made available to the collaboration. Simulation and data 
processing at the U.K. data center are performed using distributed GridPP 
resources~\cite{Faulkner:2006:gridpp,Britton:2009:gridpp}.

\subsection{The Offline Software Stack}

The LZ Offline software stack is based on standard HEP frameworks, specifically Geant4 for 
simulations~\cite{agostinelli:2002hh} and Gaudi for reconstruction~\cite{barrand:2001ny}.

The simulation package is called BACCARAT~\cite{Akerib:sims-2019}. This software provides object-oriented 
coding capability specifically tuned for noble liquid detectors, working on top of the Geant4 engine.
BACCARAT can produce detailed, accurate simulations of the LZ detector response and backgrounds,
which are crucial both for detector design and during data analysis. The physics accuracy of the LZ
simulations package was validated during the science run of LUX, as described in~\cite{Akerib:2011ec}.

BACCARAT is integrated into the broader LZ analysis framework, from production to validation and analysis. 
Two output formats are supported, a raw simulation output at the interaction level, and a reduced tree 
format at the event level. Both output files are written in the {\normalfont\ttfamily ROOT} data format~\cite{Brun:1997pa}. 
A Detector Electronics Response package can be used to emulate the signal processing done by the front-end 
electronics of LZ. It reads raw photon hits from BACCARAT to create mock digitized waveforms, organized and 
written in an identical format to the output of the LZ data acquisition system. These can be read in by 
the analysis software, providing practice data for framework development and analysis.

The data processing and reconstruction software (LZ analysis package, or LZap for short) extracts the PMT 
charge and time information from the digitized signals, applies the calibrations, looks for S1 and S2 
candidate events, performs the event reconstruction, and produces the so-called reduced quantity (RQ) files, 
which represent the primary input for the physics analyses.

LZap is based on the Hive version of the Gaudi code, which is specially designed to provide 
multi-threading at the sub-event level. The framework supports the development of different physics 
modules from LZ collaborators and automatically takes care of the basic data handling (I/O, event/run 
selection, DB interfaces, etc.). Gaudi features 
a well established mechanism for extending its input and output modules to handle custom formats. This
functionality has has been exploited in the design of the LZ raw data and RQ formats.

All non-DAQ data, i.e. any data that is not read out by the DAQ with each event, is stored in a database 
known as the ``conditions database''. This database can automatically understand the interval of validity 
for each piece of data (based on timestamps), and supports data versioning for instances such as when better 
calibrations become available. This design implements a hierarchy of data sources, which means that during 
development of code or calibrations it is possible to specify alternate sources, allowing for the validation 
of updated entries.

All LZ software is centrally maintained through a software repository based on GitLab~\cite{gitlab:web}. 
GitLab provides a continuous integration tool, allowing for 
automatic testing and installation of the offline codebase on the U.S. and U.K. data center servers. Build 
automation is inherited from the Gaudi infrastructure and supported via CMake. Release Management and Version 
Control standards were strictly enforced from a very early stage of the experiment to ensure sharing, 
verifiability and reproducibility of the results. Each code release undergoes a battery of tests before being 
deployed to production. Release management ensures that all the changes are properly communicated and documented, 
to achieve full reproducibility.

Software distribution is achieved via CernVM File System (CVMFS)~\cite{Blomer:2011zz}. CVMFS is a CERN-developed 
network file system based on HTTP and optimized to deliver experiment software in a fast, scalable, and reliable way. 
Files and file metadata are cached and downloaded on demand. CVMFS features robust error handling and 
secure access over untrusted networks~\cite{Dykstra:2014kea}. The LZ CVMFS server is 
visible to all the machines in the U.S. and U.K. data centers. All the LZ software releases and external 
packages are delivered via CVMFS: this ensures a unified data production and analysis stream, because every user can 
access identical builds of the same executables, removing possible dependencies on platform and configuration.

A rigorous program of Mock Data Challenges has been enacted, in order to validate the entire software stack and to 
prepare the collaboration for science analysis. The first data challenge simulated the initial LZ commissioning and 
tested the functionality of the reconstruction framework. The second data challenge covered the first science
run of LZ and tested the entire data analysis chain, including calibrations, detailed backgrounds and potential signals.
The third and final data challenge is currently underway (2019) and will test the complete analysis strategy, validating 
the readiness of the offline system just before the underground installation of LZ.

\section{Conclusion}
\label{sec:summary}

Considerable progress has been made towards 
implementing the LZ conceptual and technical designs 
described in Refs. ~\cite{Mount:2017qzi,akerib:2015cja}.  
The start of science operations is expected 
2020. The projected background rate enables a 1000~day exposure 
of the 5.6~tonne fiducial mass, with a spin-independent 
cross-section sensitivity 
of 1.5$\times10^{-48}$~cm$^2$ 
(90\% C.L.) at 40 GeV/c$^2$. This will 
probe a significant portion of the viable WIMP dark matter
parameter space.  
LZ is also be sensitive to spin-dependent interactions, 
through the odd neutron number isotopes $^{129}$Xe 
and $^{131}$Xe (26.4\% and 21.2\% respectively by mass).  
For spin-dependent WIMP-neutron(-proton) scattering a sensitivity 
of 2.7$\times10^{-43}$~cm$^2$ (8.1$\times10^{-42}$~cm$^2$) is expected 
at 40~GeV/c$^2$.

\section{Acknowledgements}

This work was partially supported by the U.S. Department of Energy (DOE) Office of Science under contract number DE-AC02-05CH11231 and under grant number DE-SC0019066; by the U.S. National Science Foundation (NSF); by the U.K. Science \& Technology Facilities Council under award numbers, ST/M003655/1, ST/M003981/1, ST/M003744/1, ST/M003639/1, ST/M003604/1, and \\ ST/M003469/1; and by the Portuguese Foundation for Science and Technology (FCT)under award number \\ PTDC/FIS-PAR/28567/2017; and by the Institute for Basic Science, Korea (budget number IBS-R016-D1). University College London and Lawrence Berkeley National Laboratory thank the U.K. Royal Society for travel funds under the International Exchange Scheme (IE141517). We acknowledge additional support from the Boulby Underground Laboratory in the U.K.; the University of Wisconsin for grant UW PRJ82AJ; and the GridPP Collaboration, in particular at Imperial College London. This work was partially enabled by the University College London Cosmoparticle Initiative. Futhermore, this research used resources of the National Energy Research Scientific Computing Center, a DOE Office of Science User Facility supported by the Office of Science of the U.S. Department of Energy under Contract No. DE-AC02-05CH11231. The University of Edinburgh is a charitable body, registered in Scotland, with the registration number SC005336. The research supporting this work took place in whole
or in part at the Sanford Underground Research Facility
(SURF) in Lead, South Dakota. The assistance of SURF 
and its personnel in providing physical
access and general logistical and technical support is
acknowledged. SURF is a federally sponsored research facility under Award Number DE-SC0020216.

\bibliography{LZNIM}

\end{document}